\documentclass[a4paper,10pt,twoside]{cpc-hepnp}
\usepackage[T1]{fontenc}
\usepackage{ae,aecompl}
\usepackage{longtable}

\usepackage{xcolor}
\usepackage{ulem}
\usepackage[switch]{lineno}

\usepackage{multicol}
\usepackage{graphicx}
\usepackage{booktabs}
\usepackage{amssymb,bm,mathrsfs,bbm,amscd}
\usepackage[tbtags]{amsmath}
\usepackage{lastpage}
\usepackage{enumerate}
\usepackage{multirow,amsmath,amssymb}
\usepackage{subfig}
\usepackage{slashed}
\usepackage{color}
\usepackage{verbatim}
\usepackage{caption}
\usepackage{cases}
\usepackage{hyperref}
\usepackage{float}
\usepackage{cuted}
\hypersetup{
	colorlinks=true,
	linkcolor=cyan,
	filecolor=blue,      
	urlcolor=black,
	citecolor=blue,
}

\def\mumu{\ensuremath{\mu^+\mu^-}}%

\begin{document}
\captionsetup[figure]{labelfont={bf},name={Fig.},labelsep=period}

\fancyhead[c]{\small Submitted to Chinese Physics C~~~
} \fancyfoot[C]{\small -\thepage}


\title{Expected $H \xrightarrow{} \mumu$ measurement precision with $e^{+}e^{-} \xrightarrow{} Z(q\bar{q})H$ production at the CEPC}

\author{Qi Liu$^{1,2}$\href{https://orcid.org/0000-0002-4132-5720}{\includegraphics[scale=0.05]{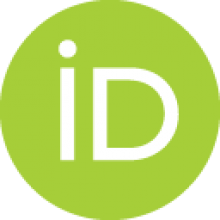}}\quad Kunlin Ran$^{2;\textcolor{blue}{1)}}$ \email {kunlin.ran@cern.ch}\quad Yanping Huang$^{2;\textcolor{blue}{2)}}$ \email {yanping.huang@cern.ch}%
\quad Gang Li$^{2}$\quad Manqi Ruan$^{2}$\quad Shan Jin$^{3}$\quad Liang Sun$^{1}$}
\maketitle

\address{%
$^1$ School of Physics and Technology, Wuhan University, Wuhan, 430072, China\\
$^2$ Institute of High Energy Physics, Beijing, 100049, China\\
$^3$ School of Physics, Nanjing University, Nanjing, 210093, China\\
}

\begin{abstract}
 A search for the dimuon decay of the Standard Model Higgs boson is performed using the Monte Carlo simulated events to mimic data corresponding to an integrated luminosity of 5.6 ab$^{-1}$ collected with the Circular Electron-Positron Collider detector  in $e^{+}e^{-}$ collisions at $\sqrt{s}=240$ GeV. The paper studies $e^{+}e^{-}\to ZH,\,Z\to q\bar{q},\,H\to \mumu$ process, and the expected significance considering only the data statistical uncertainty over the background-only hypothesis for a Higgs boson with a mass of 125 GeV is found to be
    6.1$\sigma$, corresponding to the precision of 19\%. The systematic impacts from the background Monte Carlo statistical fluctuations are estimated to be negligible. The dependence of the measurement accuracy on the muon momentum resolution of the CEPC detector has been investigated. It is found that the muon momentum resolution has to be better than 204 MeV to discover the $H\to\mu\mu$ process at the nominal integrated luminosity. And if the resolution is 100\% worse than the designed parameter, the integrated luminosity is needed to be greater than 7.2 ab$^{-1}$ to reach 5$\sigma$ significance.
\end{abstract}



\setlength{\columnsep}{1cm}
\begin{multicols}{2}

\section{Introduction}

The Standard Model (SM)~\cite{GLASHOW1961579,PhysRevLett.19.1264} of the particle physics predicts the existence of one neutral scalar particle, known as the Higgs boson~\cite{PhysRevLett.13.321, PhysRevLett.13.508, PhysRevLett.13.585}. The electro-weak spontaneous symmetry breaking is introduced by the Higgs mechanism through a complex doublet scalar field. In 2012, the ATLAS~\cite{ATLASHiggs} and CMS~\cite{CMSHiggs} Collaborations claimed the discovery of a new particle with a mass of approximately 125 GeV. Subsequent researches have manifested that this particle is consistent with the predicted Higgs boson in the SM.

The interactions between the Higgs boson and the third-generation charged fermions have been observed by both ATLAS and CMS Collaborations~\cite{Aad_2016, 2018173, Sirunyan:2018kst}. While the Higgs couplings to other generation fermions aren't observed yet. The $H\to\mu^+\mu^-$ process is important for probing the properties of the Higgs Yukawa couplings to the second generation fermions. Recently, in the ATLAS experiment, the detection significance of the $H \rightarrow \mumu$
    process is found to be 2.0$\sigma$ (1.7$\sigma$ expected) with $pp$ collision data collected at $\sqrt{s}=13$ TeV with the integrated luminosity of 139 fb$^{-1}$~\cite{2021135980}. In the CMS experiment, the significance is 3.0$\sigma$ (2.5$\sigma$ expected) with the integrated luminosity of 137 fb$^{-1}$~\cite{2021JHEP...01..148C}. In the projections with the ATLAS detector at HL-LHC (~3000 fb$^{\rm{-1}}$), the expected precision of the branching ratio (BR) of $H \rightarrow \mumu$ is
    14\%~\cite{ATL-PHYS-PUB-2018-054}. At the International Linear Collider (ILC), the combined precision of the BR ($H \rightarrow \mumu$) is estimated to be 17\%~\cite{Kawada_2020}. The relative uncertainty on the measurement of $\sigma(ZH)\times B(H\to \mu\mu)$ from the expected Future Circular Collider electron-possitron (FCC-ee) data is 19\% with the integrated luminosity of 5 ab$^{-1}$ at $\sqrt{s}=240$ GeV~\cite{Mangano:2651294}.

Like other lepton colliders, the Circular Electron Positron Collider (CEPC)~\cite{CEPC-SPPCStudyGroup:2015csa} experiment has significant advantages for Higgs boson property measurements. The signal-to-noise ratio (SNR) is significantly higher due to lepton collisions. Moreover, the Higgs boson candidates can be identified through the recoiled mass method without tagging its decay products. A previous study~\cite{Cui_2018,An_2019} was performed with the CEPC detector at the center-of-mass energy of 250 GeV with an integrated luminosity of 5 ab$^{-1}$. The detection significance from a counting experiment was reported in the $\mumu$ mass window of [124.3, 125.2] GeV. According to the new design parameters of the CEPC experiment, the center-of-mass energy would be updated to 240 GeV and the integrated luminosity would be accumulated to 5.6 ab$^{-1}$ over seven years~\cite{CEPCStudyGroup:2018ghi}. The benchmark of the detector has been optimized as well from Pre-CDR~\cite{CEPC-SPPCStudyGroup:2015csa} to CDR~\cite{CEPCStudyGroup:2018ghi}. It's interesting and important to re-study the $H\to\mu\mu$ process with the latest benchmark of the CEPC detector and the corresponding updated Monte Carlo (MC) samples. The event selections are updated and the Toolkit for Multivariate Data Analysis (TMVA) is applied to improve the sensitivity. The expected significance is estimated using the asymptotic approximation~\cite{Cowan_2011} method. Moreover, the performance of the CEPC detector is discussed by smearing the resolution of muon momentum in simulated events (Section~\ref{Discussion}).

The paper is organised as described below. Section~\ref{mcsim} briefly summarizes the CEPC detector and MC samples. Section~\ref{selection} presents object reconstruction and event selections. Section~\ref{BDTG} further optimizes the event categorization by the TMVA method. Section~\ref{Model} studies the signal and the background models. Section~\ref{sec:result} calculates the expected measurement precision of the $H\to\mu^+\mu^-$ process in the CEPC experiment. Section~\ref{Discussion} discusses the performance of CEPC detector. And Section~\ref{Conclusion} concludes the analysis.

\section{CEPC detector and MC samples}
\label{mcsim}

The baseline detector concept based on the Monte Carlo (MC) simulation studies at the CEPC is developed from the International Large detector (ILD) through a sequence of optimization~\cite{CEPCStudyGroup:2018ghi}. The detector is composed of a high precision silicon based vertex and tracking system, a Time Projection Chamber (TPC), a silicon-tungsten sampling electromagnetic calorimeter (ECAL), a resistive plate chamber (RPC)-steel sampling hadron calorimeter (HCAL), a 3-Tesla solenoid, and a muon/yoke
    system~\cite{detectorstudygroup2019cepc}. The center of mass energy ($\sqrt{s}$) of the $e^{+}e^{-}$ collision for the Higgs production is 240 GeV. A GEANT4-based detector simulation framework, MokkaPlus (an update version of Mokka~\cite{MoradeFreitas:2002kj}) is used for the CEPC detector simulation. MC events at the CEPC are generated with
the Whizard V1.9.5~\cite{Kilian:2007gr} program at leading order (LO) with initial state radiation (ISR) effects~\cite{Mo_2016} taken into account. Pythia 6~\cite{sjostrand2003pythia} is used for parton showering and hadronization with parameters tuned based on the Large Electron Positron Collider (LEP)~\cite{Taylor:2312570} data. The analysis focuses on the signal process of $e^{+}e^{-} \rightarrow Z(\to q\bar{q})H(\to\mu^{+}\mu^{-})$, where the $Z$ boson decays to 2 jets. There are two kinds of background components: the two-fermions background ($e^{+}e^{-} \rightarrow f\bar{f}$) and the four-fermions background. The four fermions in the final states can be combined into two bosons, which are $Z$ or $W$, and the processes are named as ``$ZZ$'' and ``$WW$'' respectively. Additionally, when the
    final states contain a pair of electron and the accompanying neutrino, the process will be excluded from the ``$ZZ$'' and ``$WW$'' groups and will be named as ``single $Z$'' or ``single $W$'', which indicates the origin of the left two fermions. If some final particles can come from either ``$ZZ$'' or ``$WW$'', for instance $\nu_{\mu}\bar{\nu_{\mu}}\mu^{+}\mu^{-}$, they will be named as ``$ZZ$ or $WW$ mix''. The analogous combination can also happen between the ``single $Z$'' and the ``single $W$'', which
    will be called as ``single $Z$ or single $W$''. The ``$ZZ$ or $WW$ mix'' and ``single $Z$ or single $W$'' processes are grouped as ``$Z$ or $W$'' background. For completeness, all background MC are used in the analysis. While  it can be expected that most background will be excluded after event selections (Section~\ref{selection}) in the $Z(\to q\bar{q})H(\to\mu^{+}\mu^{-})$ phase space, where there are 2 muons and 2 jets in the final states. The dominant background in the analysis is the ``$ZZ$'' process, where one of the $Z$ boson decays to 2 muons and the other one decays to 2 quarks.

    Table~\ref{tab:sample} summarizes cross sections and statistics of the MC samples used in the analysis. The signal sample is procuded with the Higgs mass at 125 GeV. The designed integrated luminosity of the collected Higgs events from the CEPC detector is 5.6 ab$^{-1}$. In order to normalize the simulated events to the expected yields of 5.6 ab$^{-1}$, scale factors are applied and shown in the table.

\begin{table*}[htbp]
\caption{Cross sections and statistics of the simulated MC samples. To normalize the simulated events to the expected yields of 5.6 ab$^{-1}$, scale factors are applied.}
    \label{tab:sample}
\centering
\begin{tabular}{cccccccc}
\hline
Process       & $Z(\to q\bar{q})H(\to\mumu)$ & Single $Z$ & Single $W$ & $WW$       & $ZZ$      & $Z$ or $W$ & 2$f$       \\ \hline
$\sigma$ [fb] & 0.02977                      & 1541.68    & 3485.25    & 9076.11    & 1140.97   & 3899.63    & 143180.71  \\
Statistics    & $\sim$100 k                  & $\sim$8 M  & $\sim$18 M & $\sim$50 M & $\sim$6 M & $\sim$20 M & $\sim$30 M \\
Norm Factor&0.0017&1.1&1.1&1.1&1.1&1.1&27\\
\hline
\end{tabular}
\end{table*}

\section{Object reconstruction and event selection}
\label{selection}

A dedicated particle flow reconstruction toolkit, ARBOR~\cite{ruan2014arbor,2018Arbor}, has been developed for the CEPC baseline detector concept~\cite{CEPCStudyGroup:2018ghi}. The matching module inside ARBOR identifies calorimeter clusters with matching tracks and builds reconstructed charged particles. In the particle flow reconstruction, muons exhibit
themselves as minimum ionizing particles in the calorimeter matched with tracks in the
tracker as well as in the muon detector. A lepton identification algorithm, LICH~\cite{2017Lich}, has been developed and implemented
in ARBOR. LICH combines discriminating variables to build lepton-likelihoods using a multivariate technique. The momenta of muons are determined by their track momenta. The particle flow algorithm provides a coherent interpretation of an entire physics event and, therefore, is well suited for the reconstruction of compound physics objects such as jets, which are formed from particles reconstructed by ARBOR using the Durham clustering algorithm ($e^{+}e^{-}$ $k_{T}$-algorithm)~\cite{CATANI1991432}. Jet energies are calibrated through a two-step process. First, calibrations are applied to particles identified by ARBOR. In the second step, the jet energy are calibrated using physics events. At the CEPC, $W$ and $Z$ bosons are copiously produced and can be identified with high efficiency and purity. Thus $W\to q\bar{q}$ and $Z\to q\bar{q}$ decays serve as standard candles for
the jet energy calibration. The enormous statistics allows the jet response to be characterized in detail. Specifically in the analysis, only muons with the momenta greater than 30 GeV are considered. Two muons with different charges are selected with the invariant mass closest to the Higgs boson mass (125 GeV). The signal process focuses on the hadronic decay of the $Z$ boson ($Z \rightarrow q\bar{q}$) due to its large branching fraction. After excluding the selected $\mu^+$ and $\mu^-$, the
    Durham algorithm reconstructs all remaining particles into two jets.

The event selections are optimized to improve the signal significance. The analysis requires at least two muons with opposite signs. The Higgs boson candidate is selected by requiring $|m_{\mu\mu}-m_{H}|<10$ GeV, where $m_{H}=125$ GeV. Due to the signature topology of 2 jets in the final state, the number of reconstructed particles should be higher than leptonic final states, which is required to be greater than 25 and less than 115. The di-jet invariant mass is close to the $Z$ boson mass, which is
    selected to be greater than 55 GeV and less than 125 GeV. The four-momentum of the $q\bar{q}\mumu$ system should be close to (0, 0, 0, $\sqrt{s}$). So the momemtum of the $q\bar{q}\mumu$ system is less than 32 GeV and the energy is greater than 195 and less than 265 GeV. To supress the contamination from the $WW$ background, the energy of the muon is required to be greater than 35 and less than 100 GeV. The momenta of the missing energies are required to be less than 20 GeV in both the $x$- and $y$-axis. And the solid angle between the $q\bar{q}\mu$ system
    and the other muon is required to be greater than 2.5 rad. To supress the contaminations from the hadronic backgrounds, the momentum of the di-muon system is required to be greater than 18 GeV and less than 72 GeV. 

\begin{figure*}[htbp]
\centering
\begin{minipage}{.46\textwidth}
  \centering
  \includegraphics[width=\linewidth]{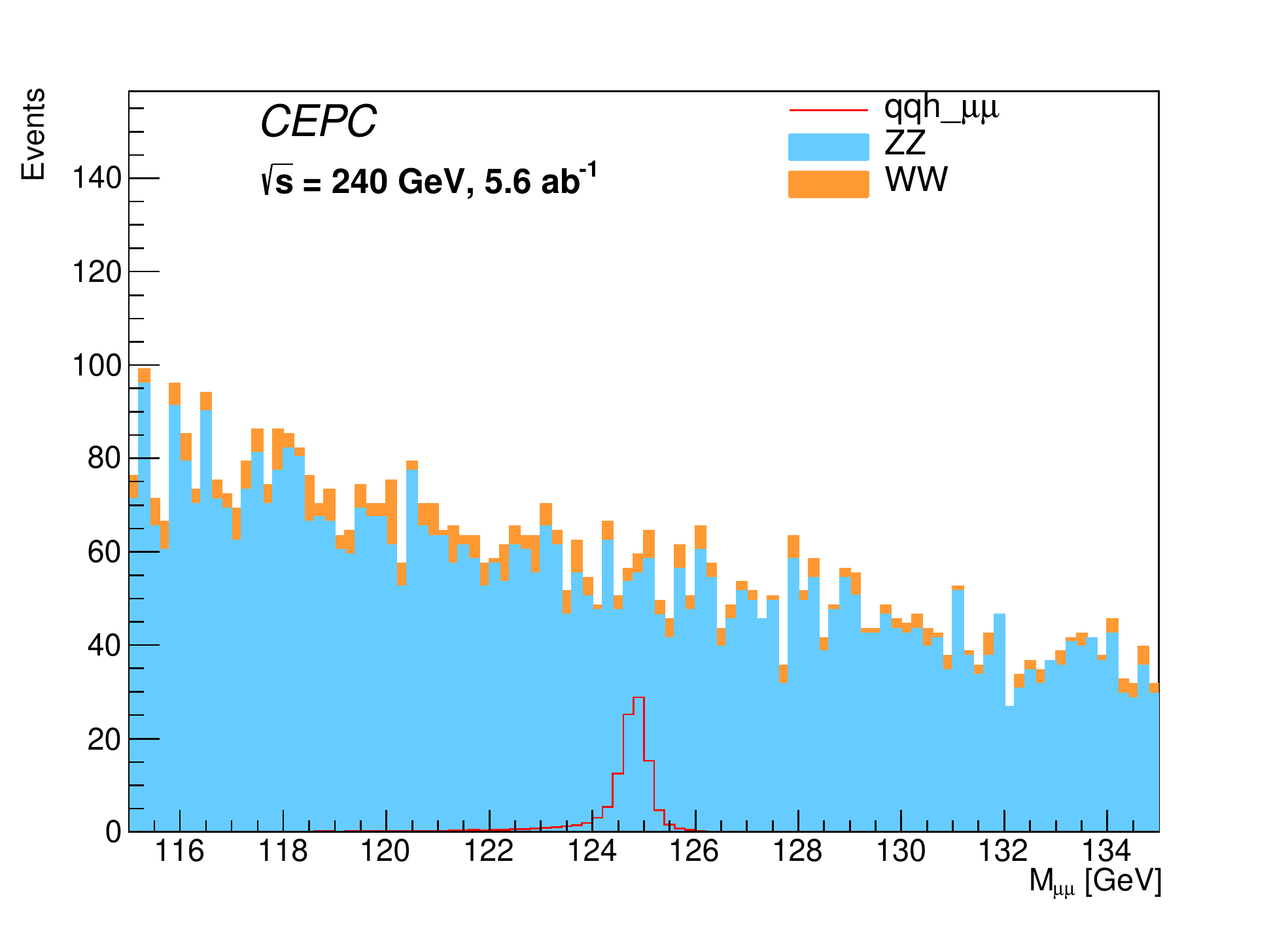}
  \captionof{figure}{The $\mumu$ invariant mass distributions of the signal and background events after the event selections. The red curve is the $Z(\to q\bar{q})H(\to\mumu)$ signal. The azure histogram is the $ZZ$ background and the orange histogram is the $WW$ background.}
  \label{fig:m_mumu}
\end{minipage}%
\hspace{1cm}
\begin{minipage}{.46\textwidth}
  \centering
  \includegraphics[width=\linewidth]{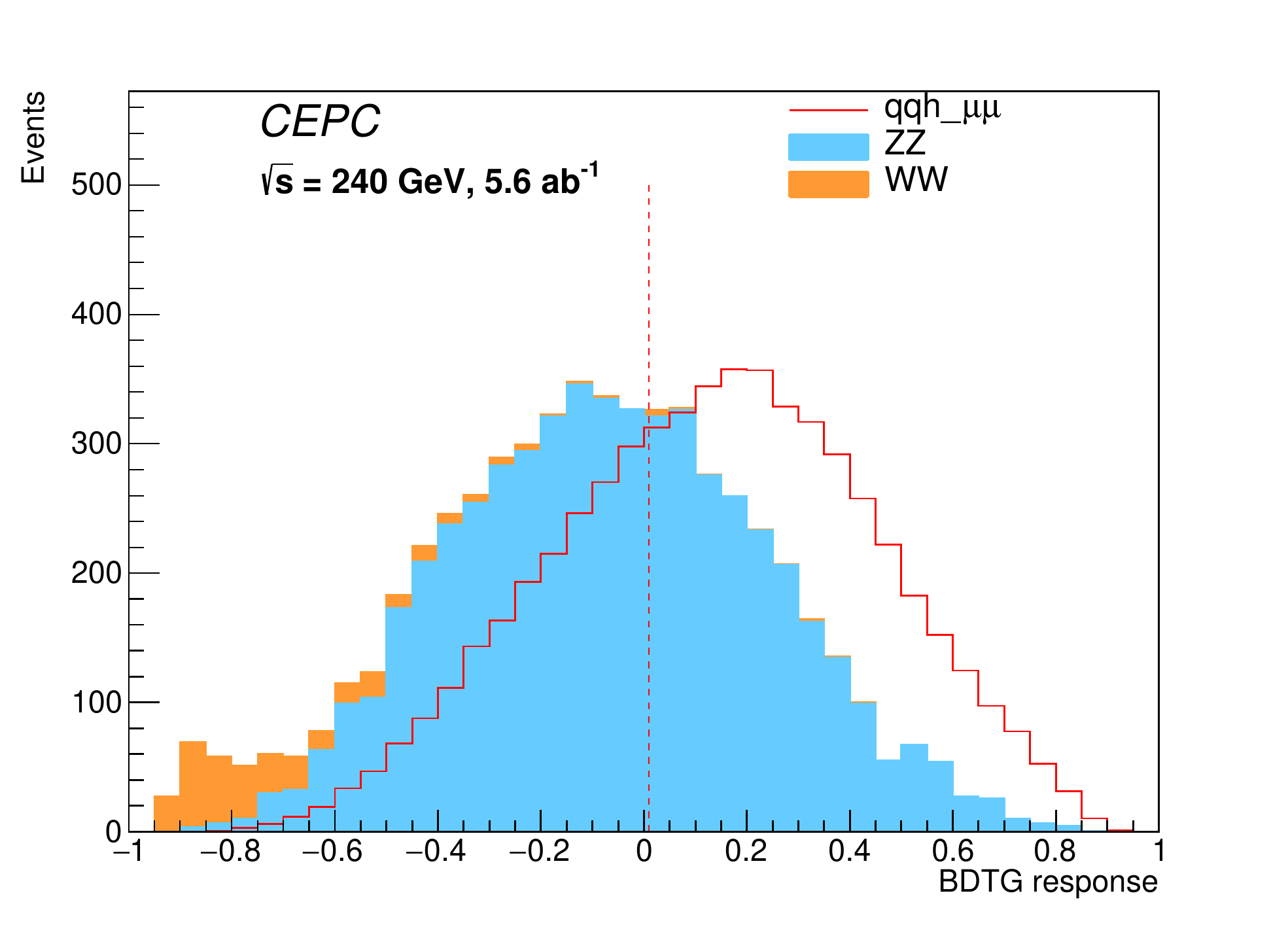}
  \captionof{figure}{BDTG reponse after training. The red curve is the $Z(\to q\bar{q})H(\to\mumu)$ signal. The azure histogram is the $ZZ$ background and the orange histogram is the $WW$ background. The signal yield is scaled to the background yield.}
  \label{fig:BDTG}
\end{minipage}
\end{figure*}

    After the selections, the signal efficiency is 77\%. The dominant background is the $ZZ$ process decaying to di-muon and di-jet, of which the fraction is 93\% in the total background. The remaining background is the $WW$ process. Contributions from other background processes are found to be negligible. The signal region is defined as $115<m_{\mumu}<135$ GeV. Figure~\ref{fig:m_mumu} shows the $\mumu$ invariant mass distributions of the signal and background events after the event selections. The red curve is the $Z(\to q\bar{q})H(\to\mumu)$ signal. The azure histogram is the $ZZ$ background and the orange
    histogram is the $WW$ background. The signal detection significance through a counting experiment (counting significance) is defined as $Z=\sqrt{2[(S+B)\ln(1+\frac{S}{B})-S]}$, where $S$ and $B$ are the corresponding signal and background yields in the $\mumu$ mass region of [124.1, 125.5] GeV, which is triple the resolution of the signal model fitted by the Double Sided Crystal Ball function (DSCB, see Section~\ref{Model}). The significance is estimated to be 4.9$\sigma$.

\section{Event categorization}
\label{BDTG}

\begin{figure*}[htbp]
     \centering
     \subfloat[$\cos\theta_{q\bar{q}}$]{\includegraphics[width=0.33\linewidth]{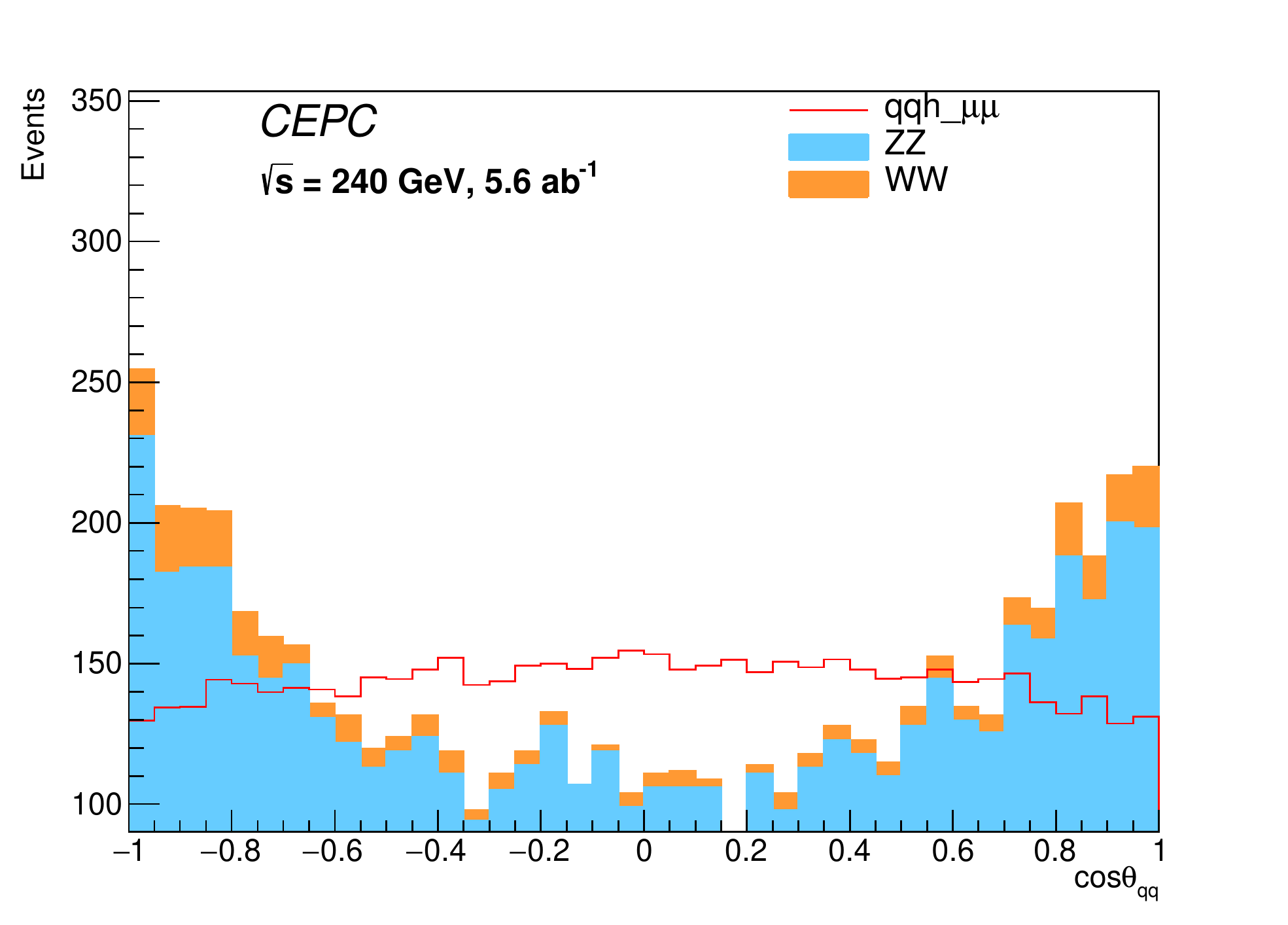}}
     \subfloat[$m_{q\bar{q}}$]{\includegraphics[width=0.33\linewidth]{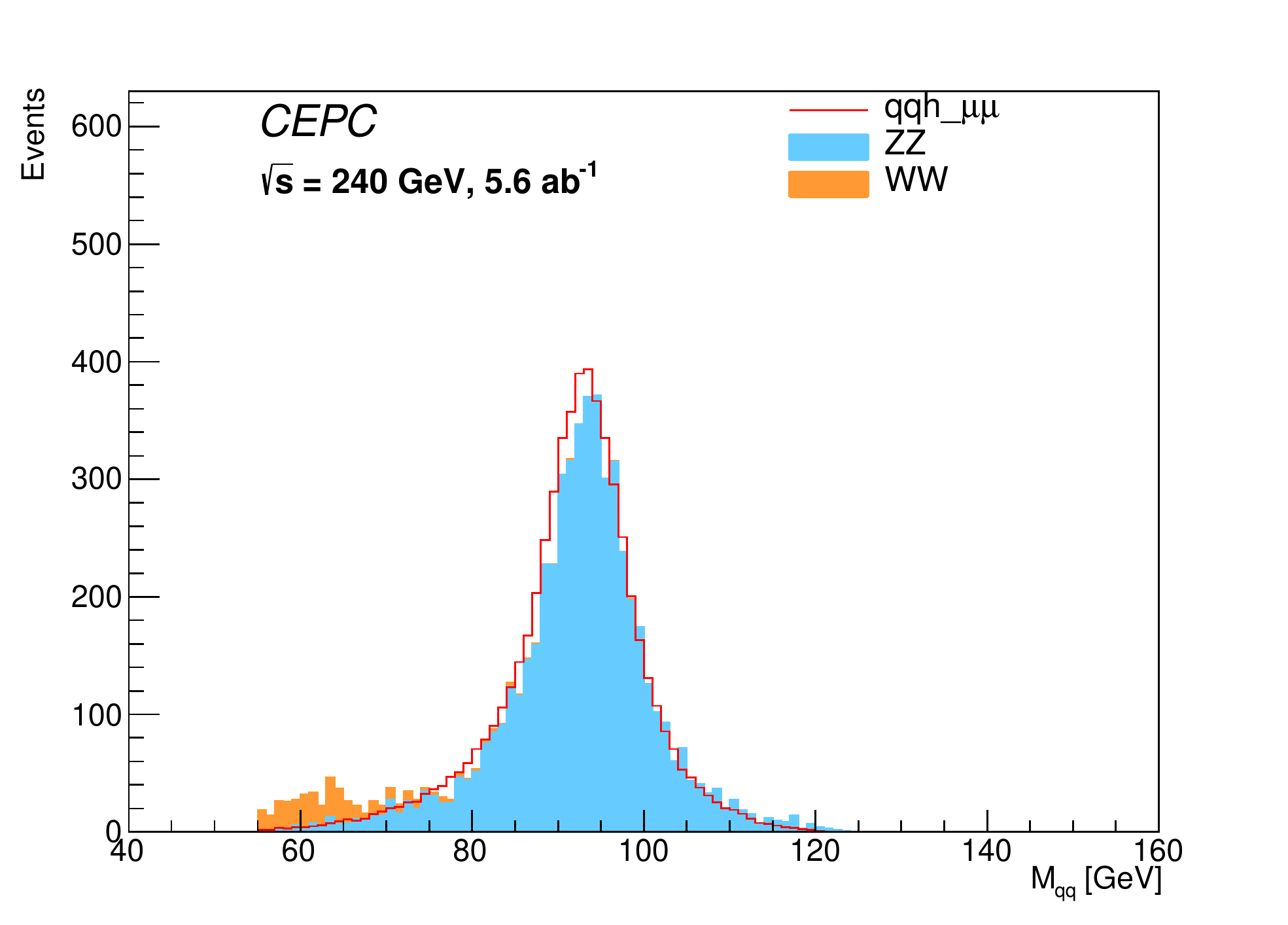}}
     \subfloat[$\Delta_{q2,\mu^+}$]{\includegraphics[width=0.33\linewidth]{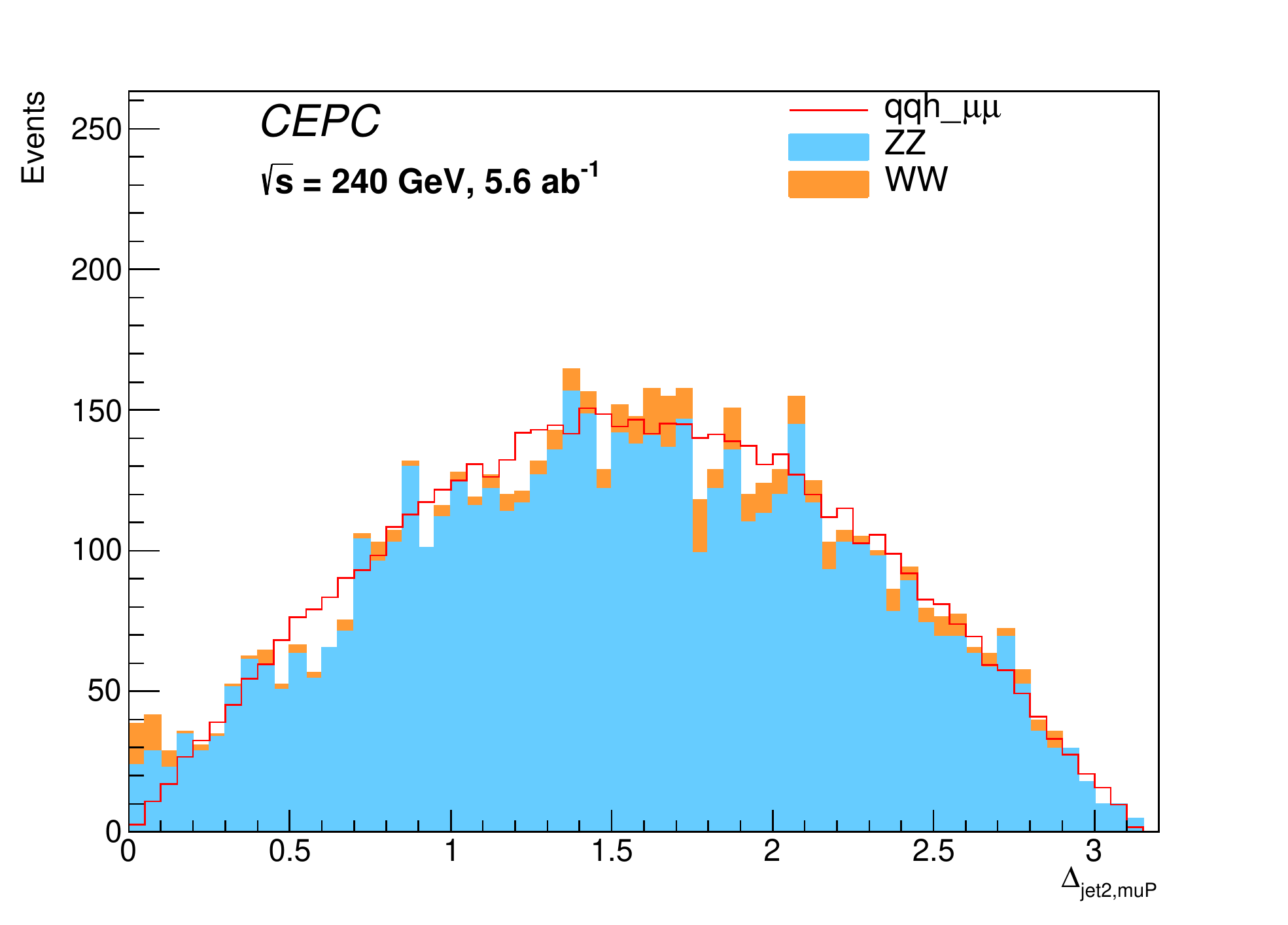}}\\
     \subfloat[$\Delta_{q1,\mu^-}$]{\includegraphics[width=0.33\linewidth]{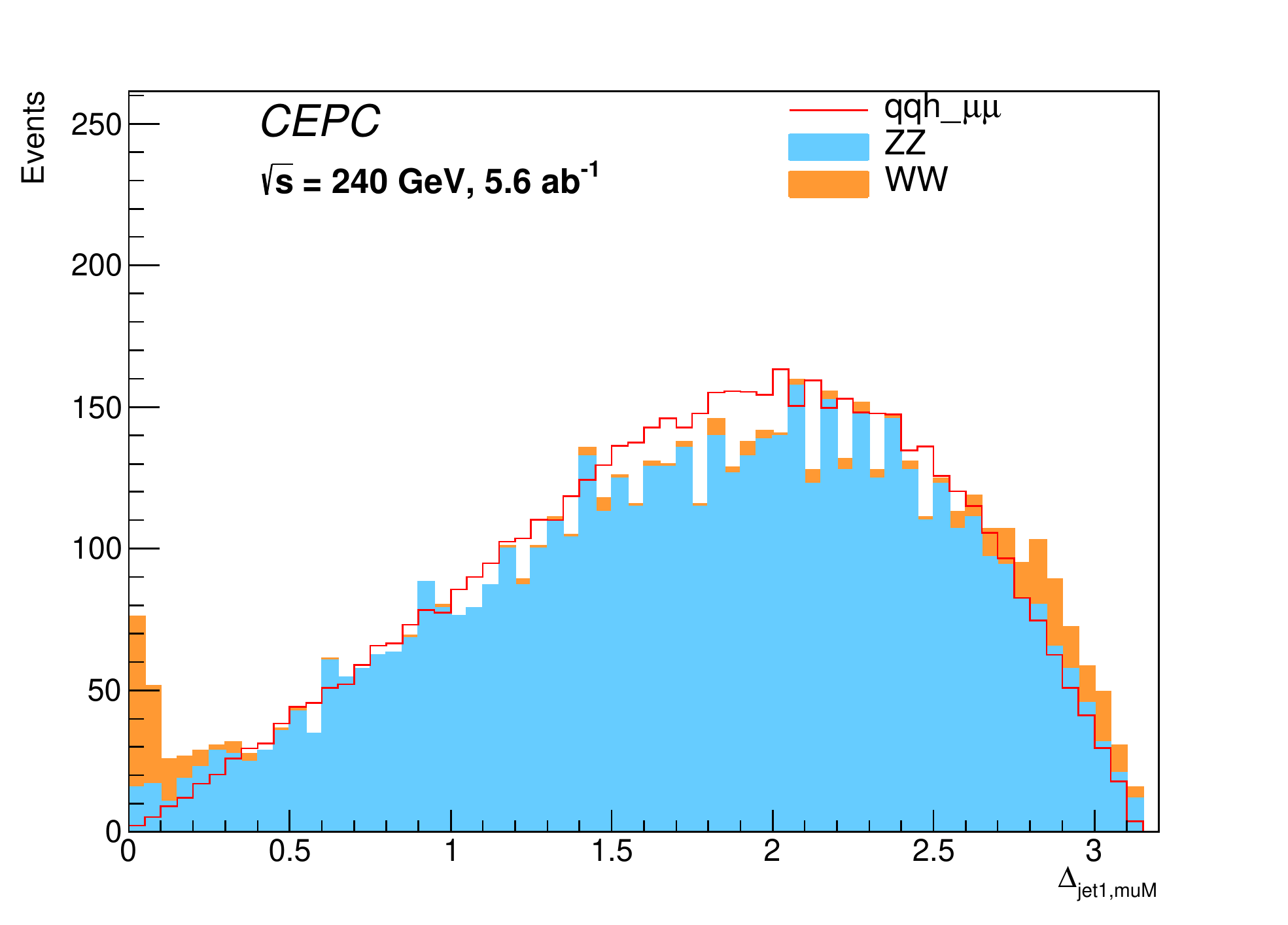}}
     \subfloat[$p_{x}^{q2}$]{\includegraphics[width=0.33\linewidth]{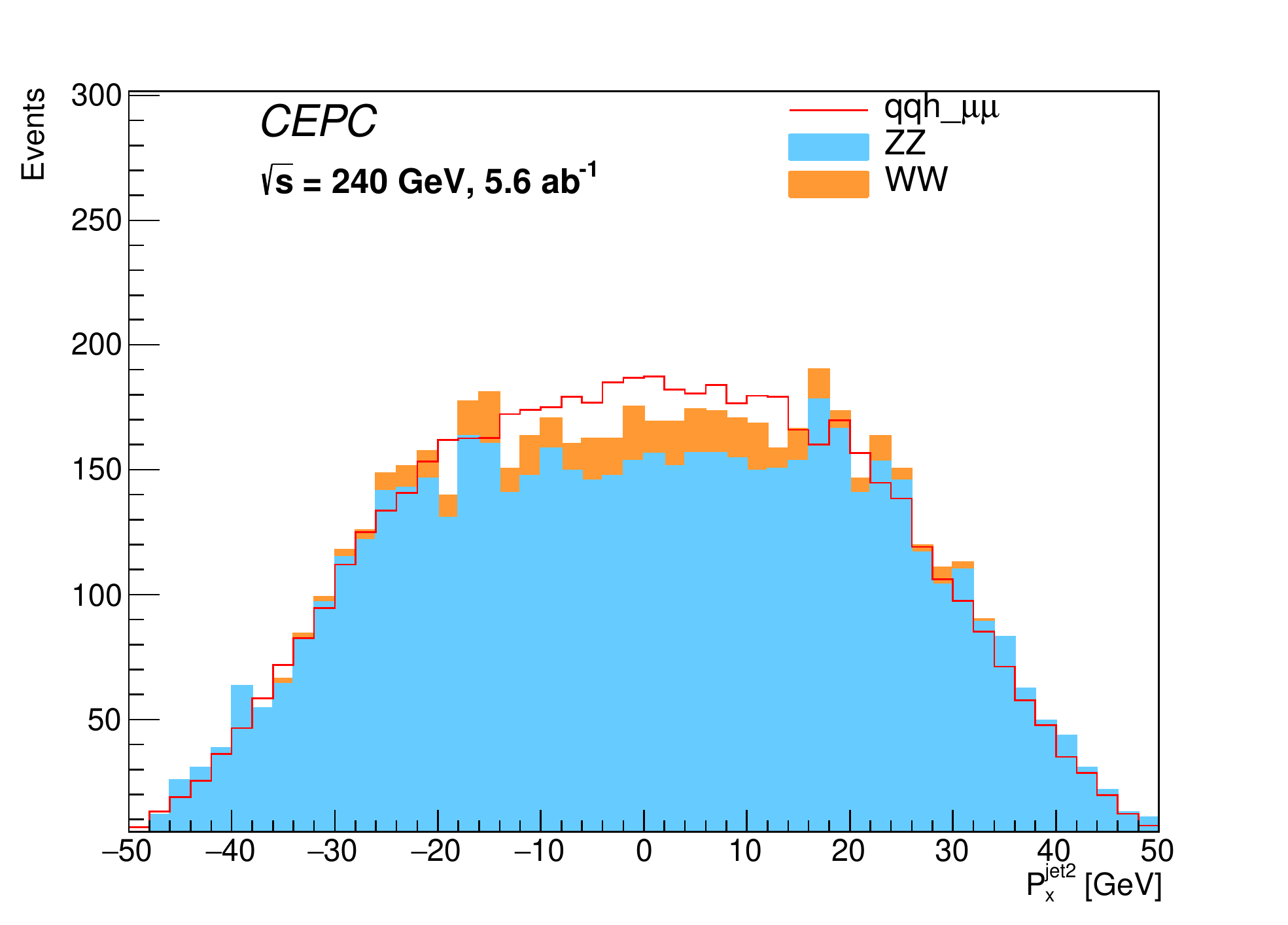}}
     \subfloat[$p_{y}^{q2}$]{\includegraphics[width=0.33\linewidth]{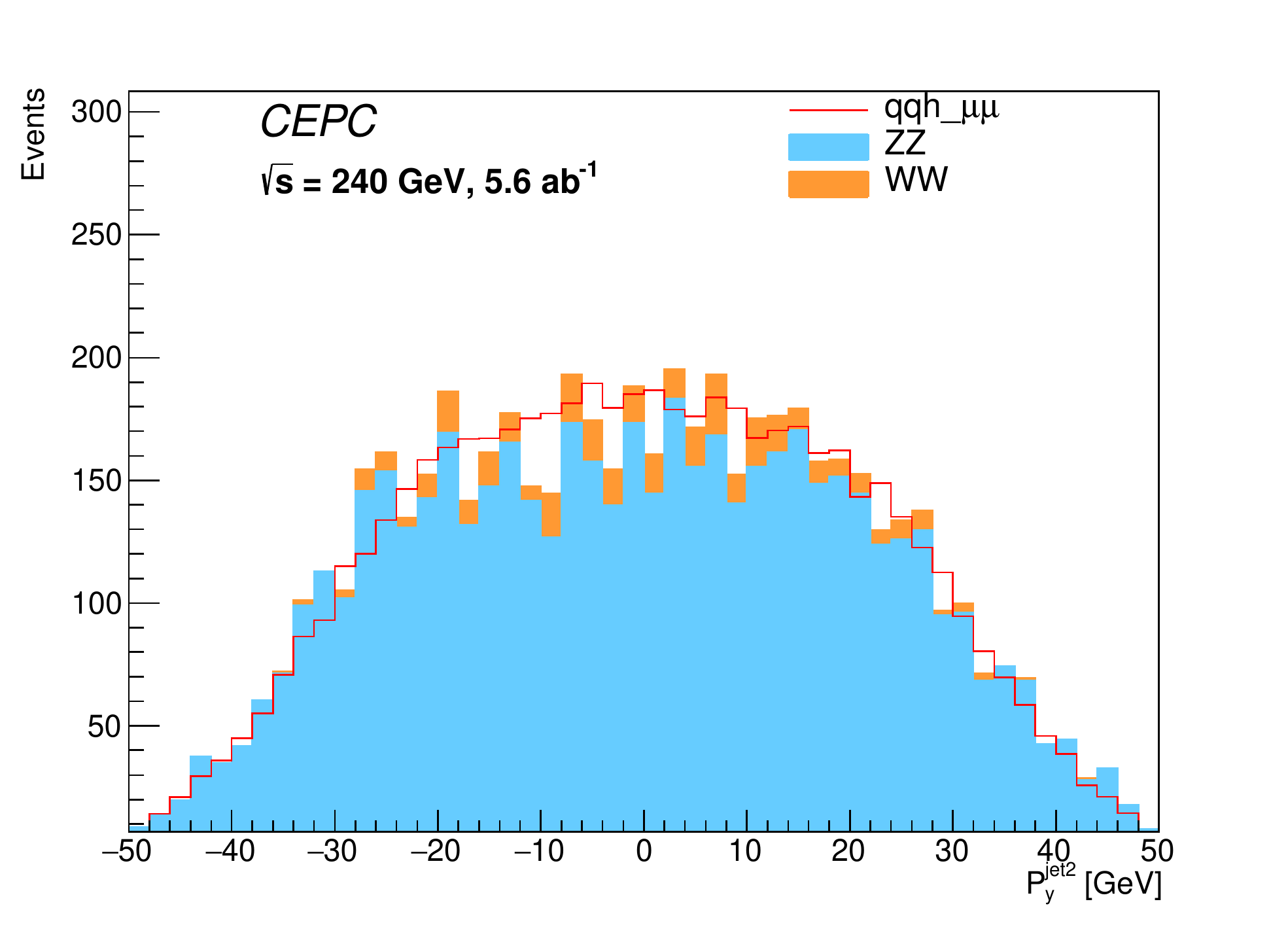}}\\
     \subfloat[$\Delta_{\mu^-,\mu^+}$]{\includegraphics[width=0.33\linewidth]{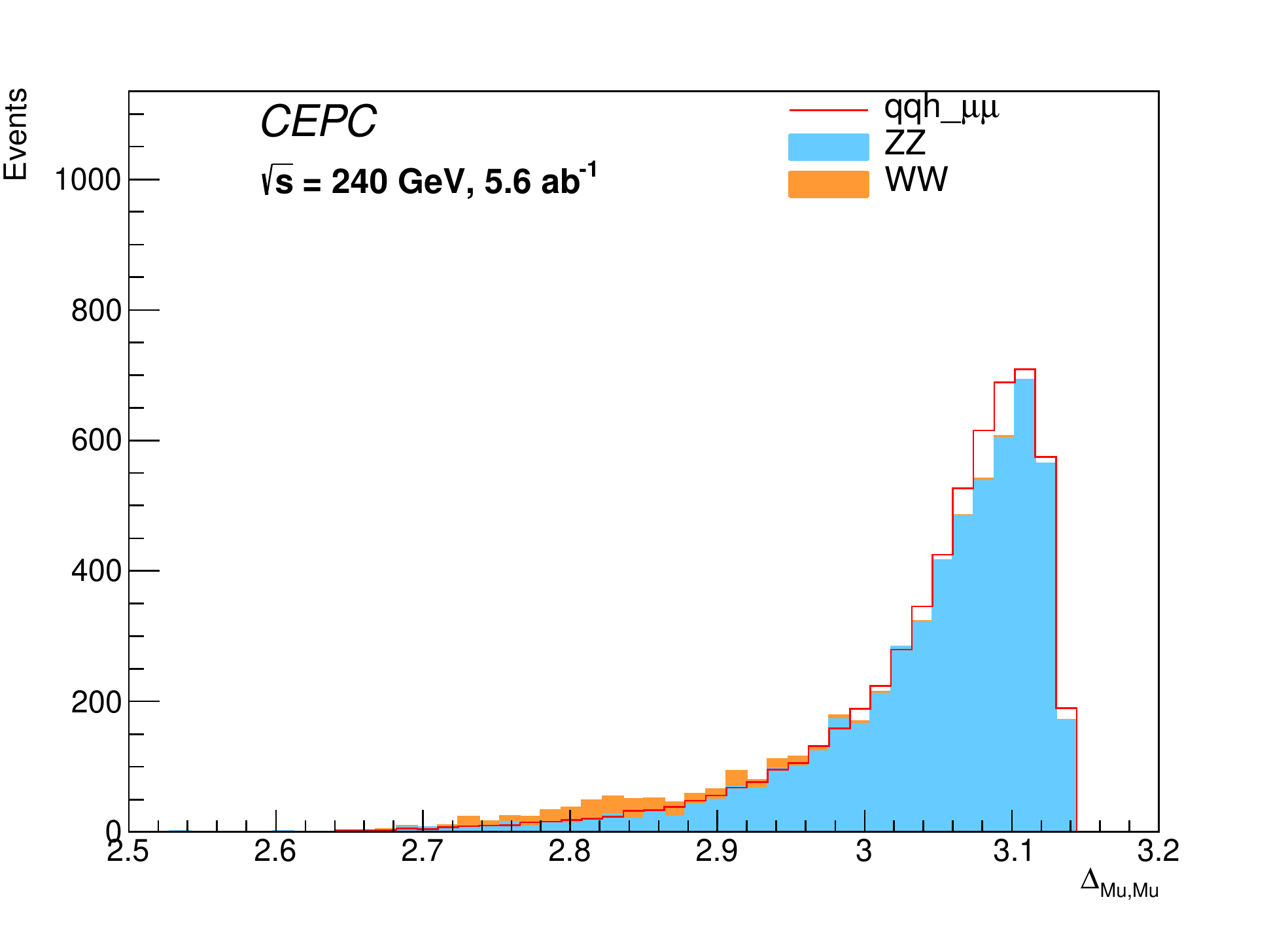}}
     \subfloat[$\cos\theta_{q2}$]{\includegraphics[width=0.33\linewidth]{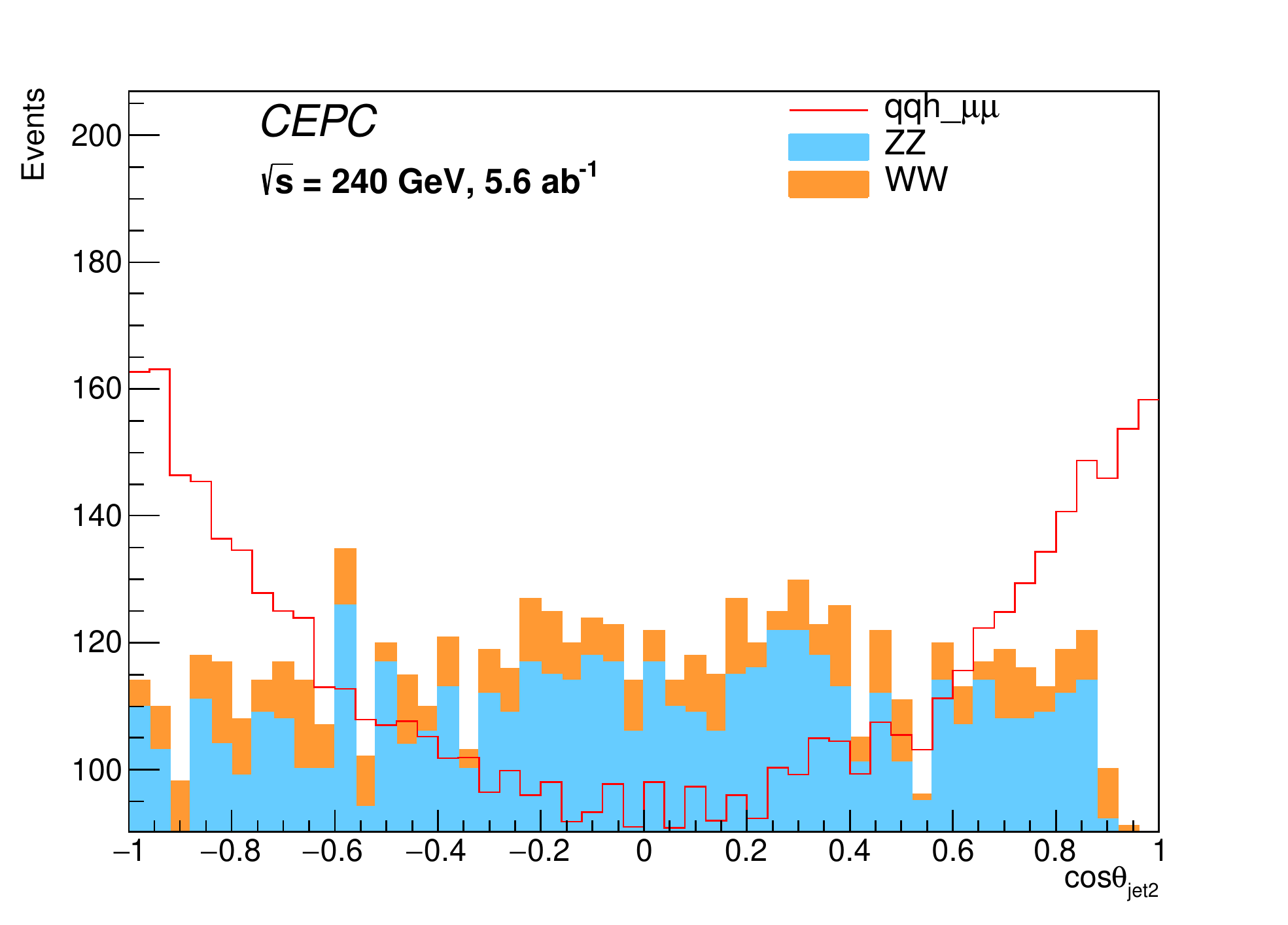}}
     \subfloat[$\cos\theta^{*}_{\mu^{+},\mu^{-}}$]{\includegraphics[width=0.33\linewidth]{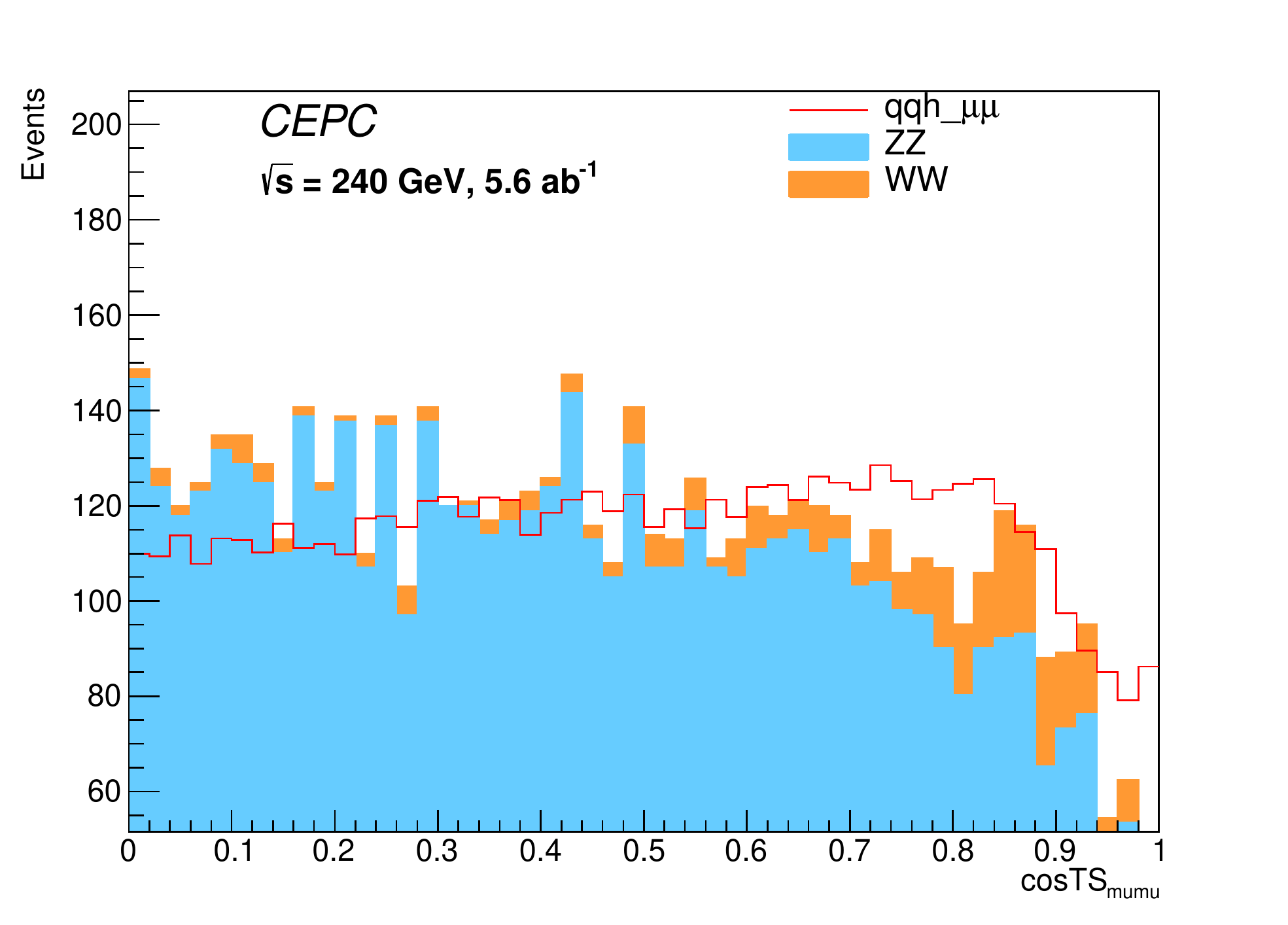}}
     \figcaption{The signal and background distributions of nine discriminant variables. The red curve is the $Z(\to q\bar{q})H(\to\mumu)$ signal. The azure histogram is the $ZZ$ background and the orange
histogram is the $WW$ background. Backgrounds are normalized to the corresponding cross sections times the integrated luminosity. The signal yield is scaled to the background yield.}
     \label{fig:trainVar}
\end{figure*}

TMVA technology is applied to categroize events for further optimizing the signal significance. Gradient Boosted Decision Trees (BDTG) method is used in the analysis. After event selections, nine discriminant variables are used for the Multivariate Data Analysis (MVA) training to separate the signal and the background processes: $\cos\theta_{q\bar{q}}$, $m_{q\bar{q}}$, $\Delta_{q2,\mu^+}$ ($\Delta$ means the solid angle, $q1/q2$ means the leading/sub-leading jet), $\Delta_{q1,\mu^-}$, $p_{x}^{q2}$, $p_{y}^{q2}$, $\Delta_{\mu^-,\mu^+}$, $\cos\theta_{q2}$ and
$\cos\theta^{*}_{\mu^{+},\mu^{-}}$\footnote{$\cos\theta^{*}_{\mu^{+},\mu^{-}}=\frac{(E_{\mu^+} + p_{z}^{\mu^+}) \times (E_{\mu^-} - P_{z}^{\mu^-}) - (E_{\mu^+} - P_{z}^{\mu^+}) \times (E_{\mu^-} + P_{z}^{\mu^-})}{m_{\mumu} \times \sqrt{m_{\mumu}^2 +  p_{\mumu}^2}}$}. The signal and background distributions of those variables are performed in the Figure~\ref{fig:trainVar}. The red curve is the $Z(\to q\bar{q})H(\to\mumu)$ signal. The azure histogram is the $ZZ$ background and the orange
histogram is the $WW$ background. In the Figure~\ref{fig:trainVar}, backgrounds are normalized to the corresponding cross sections times the integrated luminosity accounting for selection efficiencies. The signal yield is scaled to the total background yield.

\begin{figure*}[htbp]
     \centering
     \subfloat[Tight]{\includegraphics[width=0.5\linewidth]{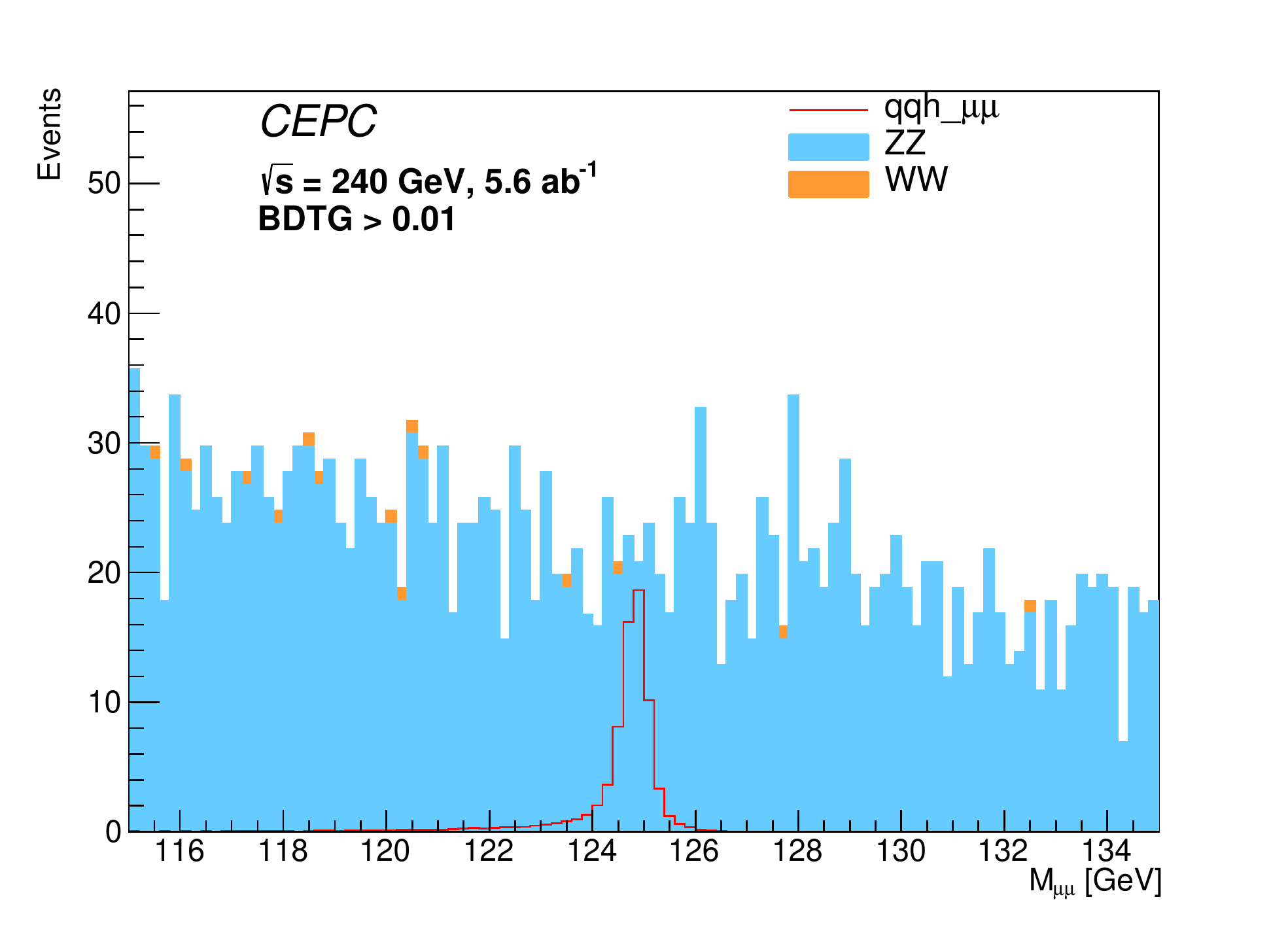}}
     \subfloat[Loose]{\includegraphics[width=0.5\linewidth]{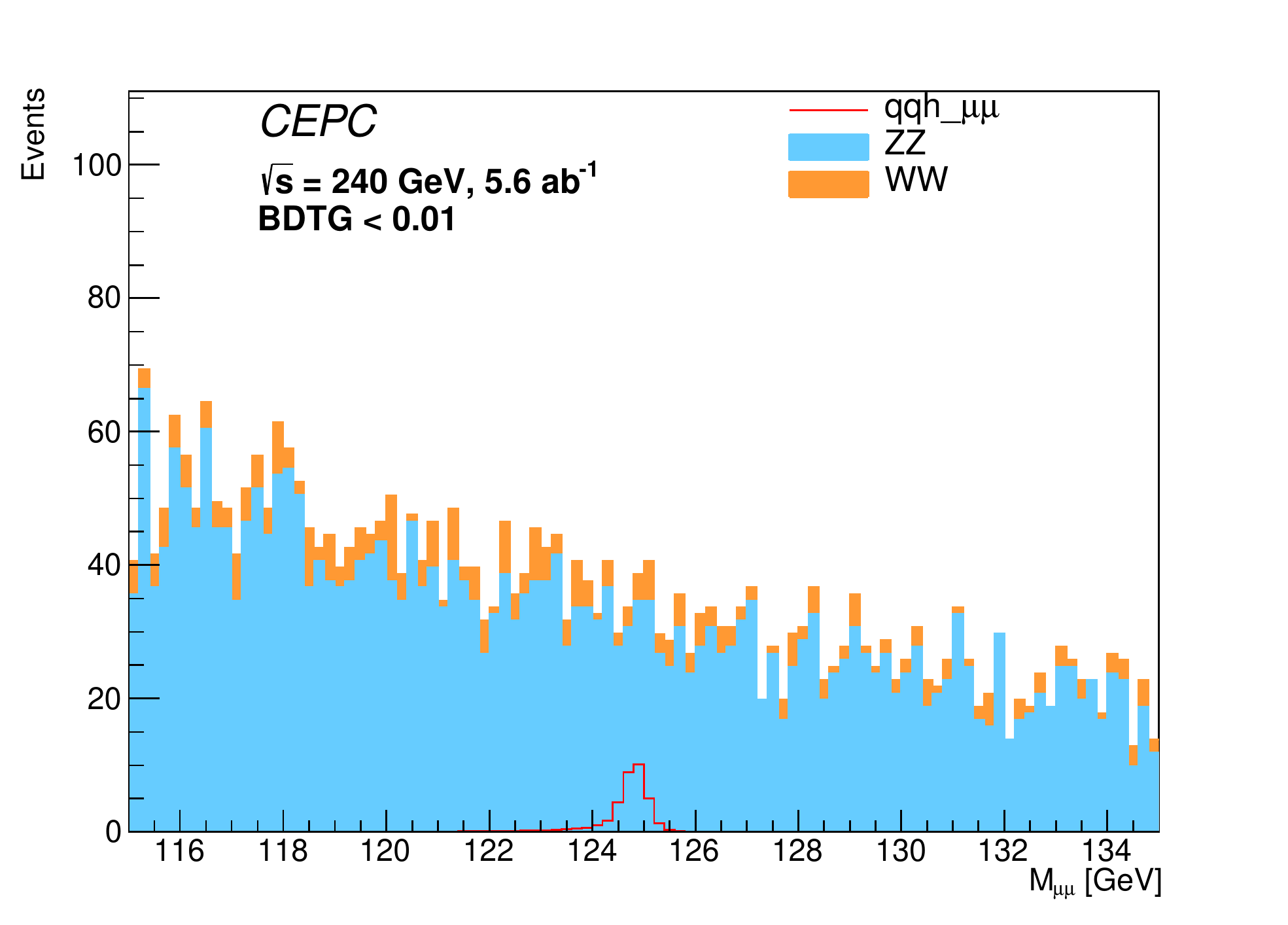}}
     \caption{$m_{\mumu}$ distributions in the tight (a) and loose (b) categories. The azure histogram is the $ZZ$ background and the orange histogram is the $WW$ background.}
     \label{fig:m_mumu_cate}
\end{figure*}

Events are equally separated into the training and the test subsets. The training events are trained by the BDTG method to classify the signal and background with an output discriminating variable constructed from nine input variables. To reduce the potential over-training effects, only test events are used to evaluate the goodness of the signal and background classification. The BDTG distribution of total events is shown in the Figure~\ref{fig:BDTG}. The signal significance is estimated as a function of BDTG response to find the optimal cut to classify 2 event categories, where the greatest total counting significance ($Z_{total}=\sqrt{Z_{1}^{2}+Z_{2}^{2}}$)
is obtained. As the result, tight (BDTG > 0.01) and loose (BDTG < 0.01) categories are defined. Figure~\ref{fig:m_mumu_cate} shows the $m_{\mumu}$ distributions in the tight (a) and loose (b) categories. The combined counting significance is estimated to be 5.6$\sigma$, with 14\% improvement with respect to the inclusive significance (4.9$\sigma$). The tight category contributes most to the sensitivity with a significance of 5.2$\sigma$. The event yields of the signal and background components in each category are summarized in the
Table~\ref{tab:event}, where signal and background yields are normalized to the corresponding cross sections times the integrated luminosity of 5.6 ab$^{-1}$.

\begin{center}
\tabcaption{The event yields of the signal and background components in each category.}
\begin{tabular}{ccccc}
     \hline
    Category& $Z(\to q\bar{q})H(\to\mumu)$ &$WW$ &$ZZ$\\ \hline
    Tight& 84.50& 16& 2461\\
    Loose& 44.25& 386& 3590\\\hline
    Total &128.75& 402& 6051 \\ \hline
\end{tabular}
\label{tab:event}
\end{center}

\section{Signal and background models}
\label{Model}

The observable of the analysis is $m_{\mumu}$ since the di-muon final state can be fully reconstructed with excellent efficiency. The narrow resonance rising above a smooth background in the $m_{\mumu}$ distribution can be used to extract the Higgs boson signal with good mass resolution. The signal model is described by the Double Sided Crystal Ball (DSCB) function:

\footnotesize
\begin{equation}\label{eq:dSCB}
f(t)=N\times
\begin{cases}
        e^{-\frac{1}{2}t^{2}},& -\alpha_{L}\le t\le\alpha_{H}\\
        e^{-\frac{1}{2}\alpha_{L}^{2}}\left[\frac{\alpha_{L}}{n_{L}}\left(\frac{n_{L}}{\alpha_{L}}-\alpha_{L}-t\right)\right]^{-n_{L}},& t<-\alpha_{L}\\
        e^{-\frac{1}{2}\alpha_{H}^{2}}\left[\frac{\alpha_{H}}{n_{H}}\left(\frac{n_{H}}{\alpha_{H}}-\alpha_{H}+t\right)\right]^{-n_{H}},& t>\alpha_{H}.
\end{cases}
\end{equation}
\normalsize

\begin{figure*}[htbp]
     \centering
     \subfloat[Tight]{\includegraphics[width=0.5\linewidth]{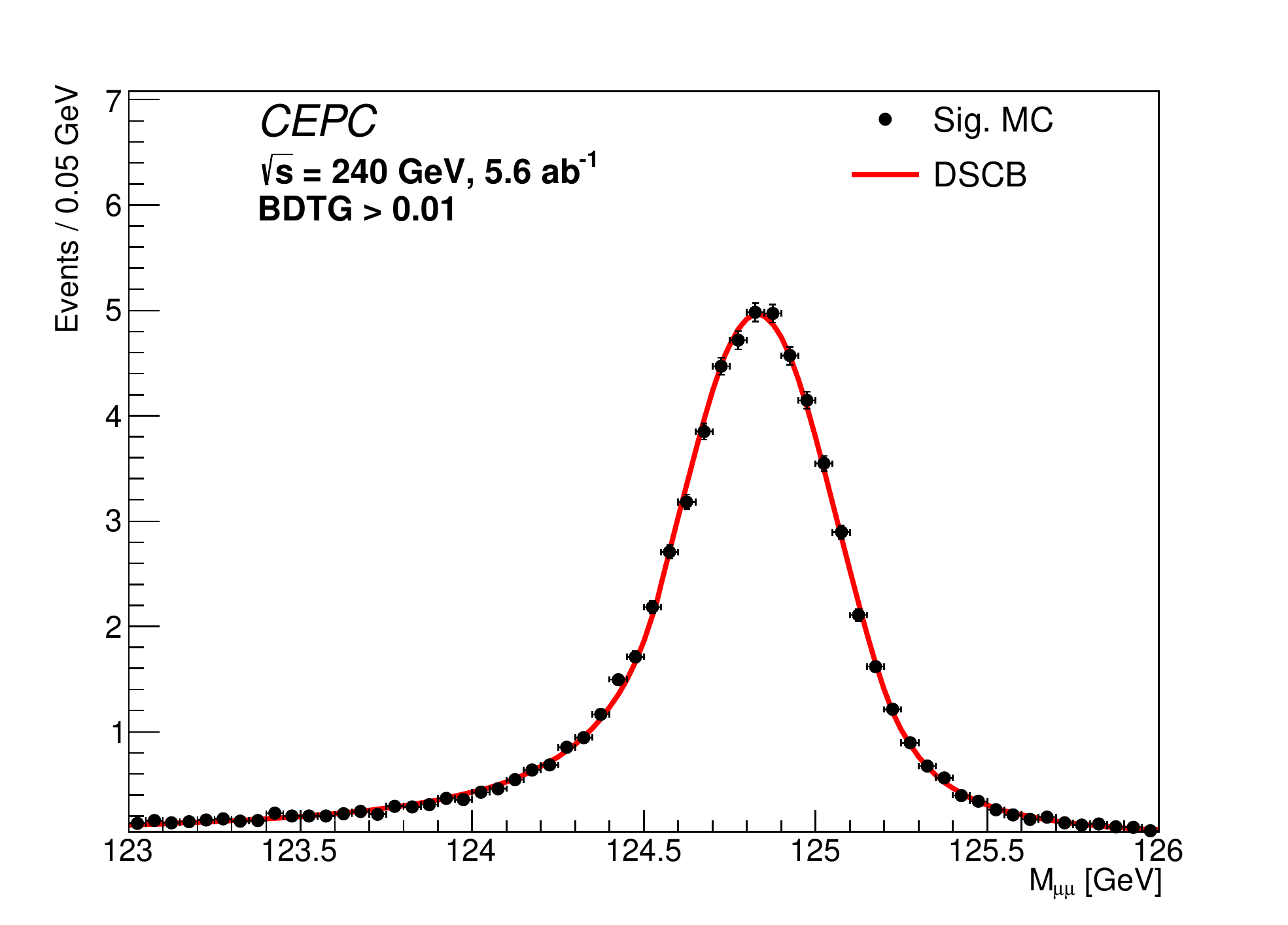}}
     \subfloat[Loose]{\includegraphics[width=0.5\linewidth]{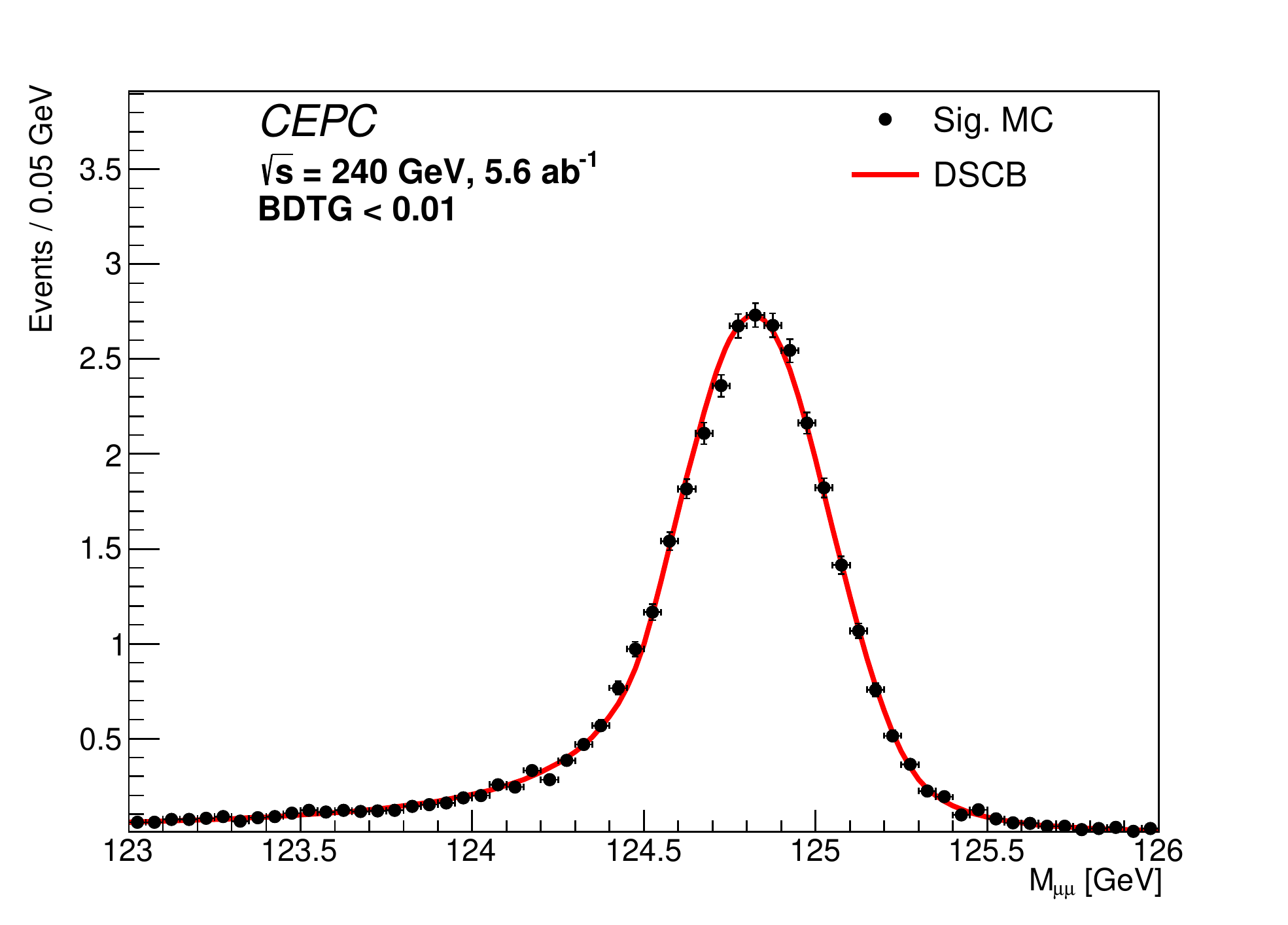}}
     \caption{The $m_{\mumu}$ distribution of the signal process and the fitted DSCB curve in the tight (a) category and the loose (b) category.}
     \label{fig:sigModel}
\end{figure*}

Where $t=(m_{\mumu}-\mu_{CB})/\sigma_{CB}$. Figure~\ref{fig:sigModel} shows the $m_{\mumu}$ distributions of the signal process and the fitted DSCB curves in 2 categories. The DSCB can describe the signal $m_{\mumu}$ distribution very well. The $\mu_{CB}$ is estimated to be $124.83$ GeV ($124.82$ GeV) in the tight (loose) category. The resolution ($\sigma_{CB}$) is estimated to be $0.23$ GeV ($0.22$
GeV) for the tight (loose) category.

%
%

Several background functions (Cheybychev polynomials, exponential functions, polynomials, etc.) are employed to fit the background mass distributions and the second-order Chebychev function is selected in the end due to the minimum $\chi^{2}$ obtained in the fits. The function is described as:

\begin{equation}\label{eq:bkgModel}
    f(m_{\mumu})=N\times[1+a_{0}m_{\mumu}+a_{1}(2m_{\mumu}^{2}-1)].
\end{equation}

\begin{figure*}[htbp]
     \centering
     \subfloat[Tight]{\includegraphics[width=0.5\linewidth]{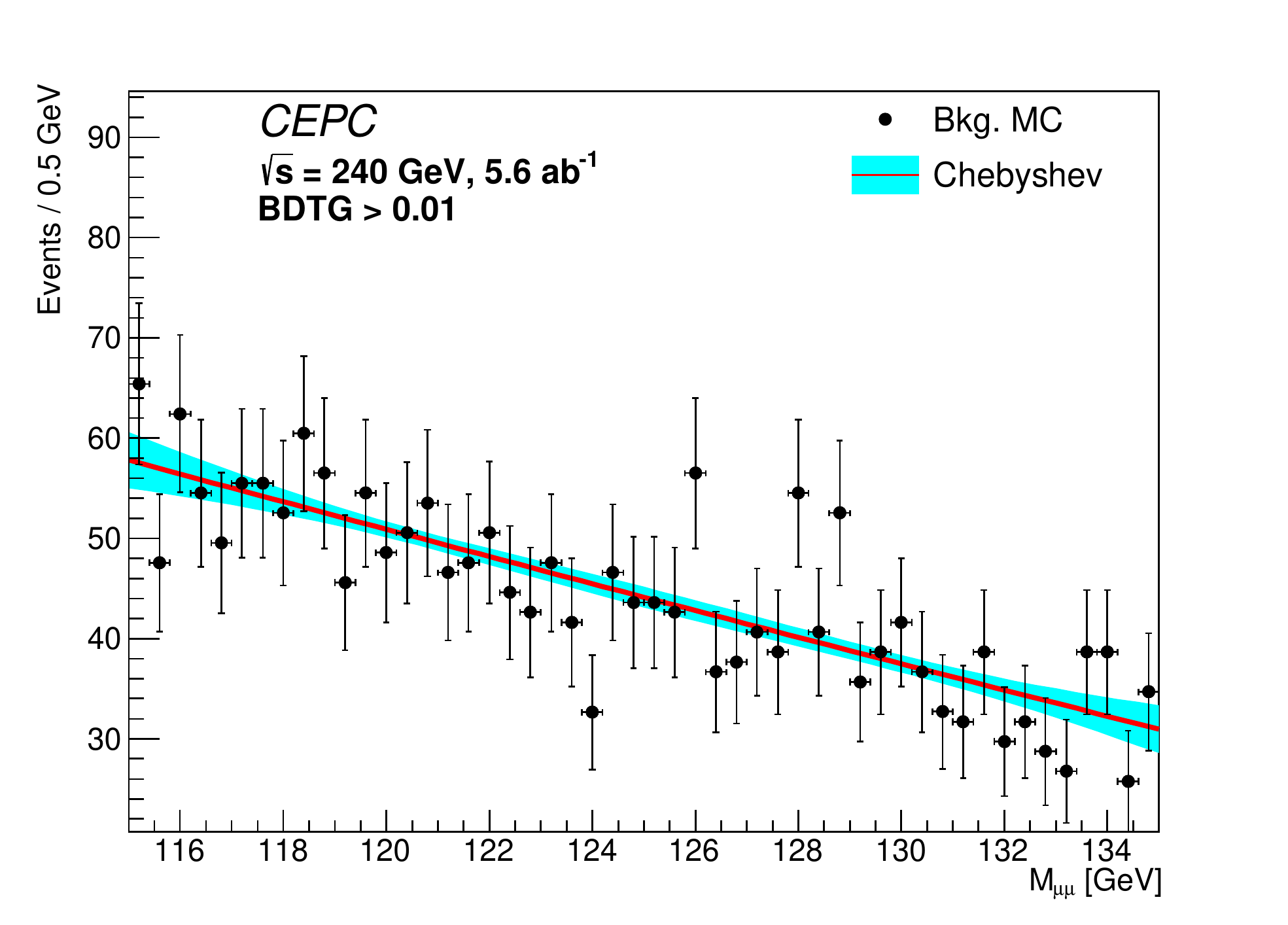}}
     \subfloat[Loose]{\includegraphics[width=0.5\linewidth]{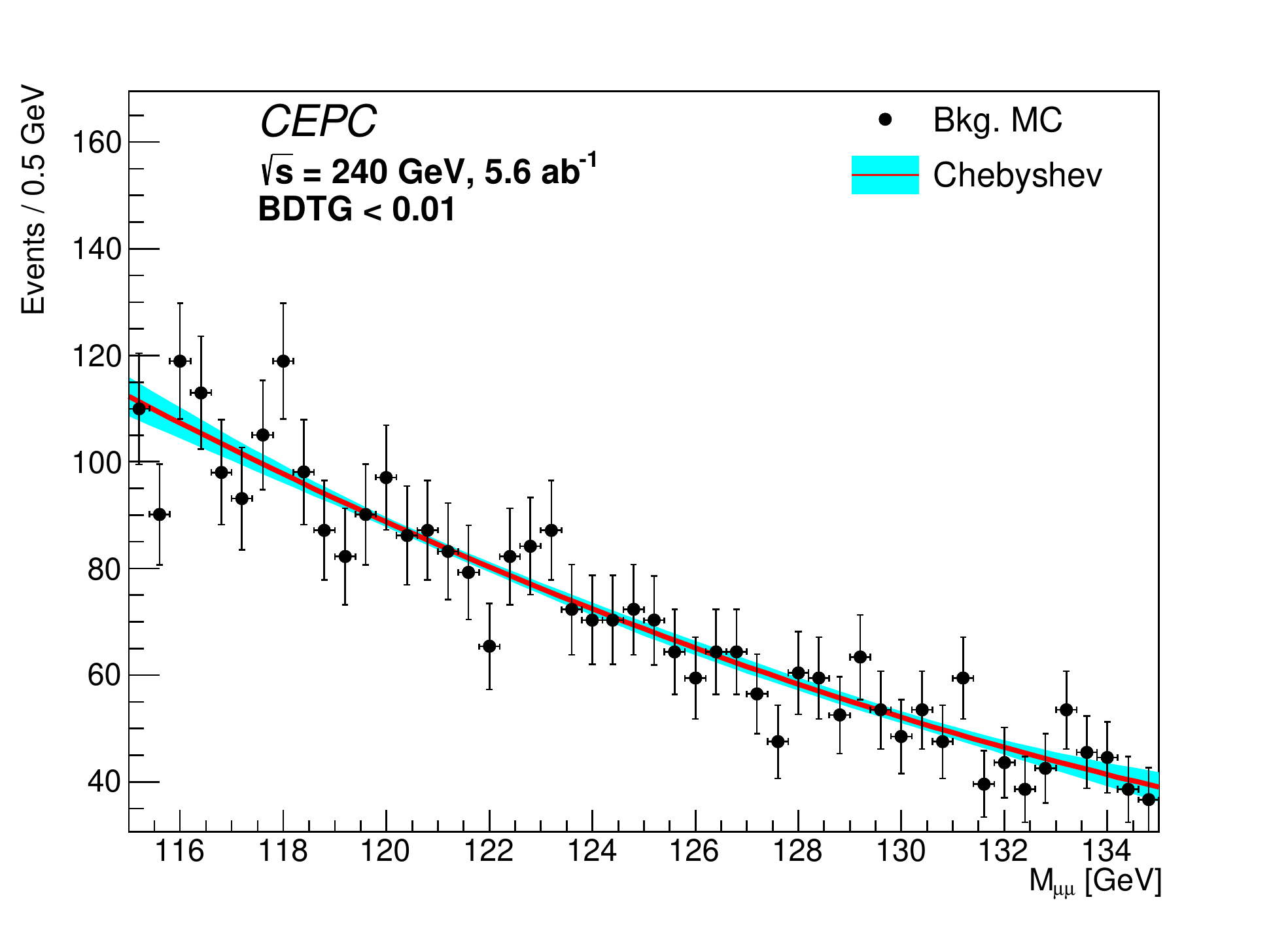}}
     \caption{The $m_{\mumu}$ distribution of the background processes and the fitted result in the tight (a) category and the loose (b) category.}
     \label{fig:backgroundModel}
\end{figure*}

Figure~\ref{fig:backgroundModel} shows the background MC mass distributions and the fitted results in two categories.

\section{Results}
\label{sec:result}

In the statistical analysis, pseudo-data is employed to mimic the real $m_{\mumu}$ distribution of the observed data collected by the CEPC detector, which is constructed by combining the signal and background MC events. The expected signal events is extracted from the pseudo-data by fitting on the $m_{\mumu}$ distribution in 2 categories simultaneously. The unbinned maximum likelihood method is used and
the fitting range is [115, 135] GeV. The likelihood function is defined as:

\begin{multline}
    \mathcal{L}\left(m_{\mumu}\right)=\prod_{c}\bigg(\operatorname{Pois}(N \mid \mu S+B)\cdot\\
    \prod_{n=1}^{N} \frac{\mu S \times f_{S}\left(m_{\mumu}\right)+B \times f_{B}\left(m_{\mumu}\right)}{\mu S+B}\bigg).
\end{multline}

Where $N$ is the pseudo event number in the category $c$, the signal strength is defined as the ratio of the measured signal yield to the one expected in the SM: $\mu=\frac{N(e^{+}e^{-}\to Z(\to q\bar{q})H(\to\mumu))}{N^{SM}(e^{+}e^{-}\to Z(\to q\bar{q})H(\to\mumu))}$, which is the parameter of interest (POI) in the analysis. $S$ and $B$ are the expected signal and background events in the category $c$, $f_{S}$ and $f_{B}$ are the signal and background models in the category $c$. In the fitting, signal model parameters are fixed to those in fitting
the signal MC and the background model parameters are floated.

\begin{figure*}[htbp]
     \centering
     \subfloat[Tight]{\includegraphics[width=0.5\linewidth]{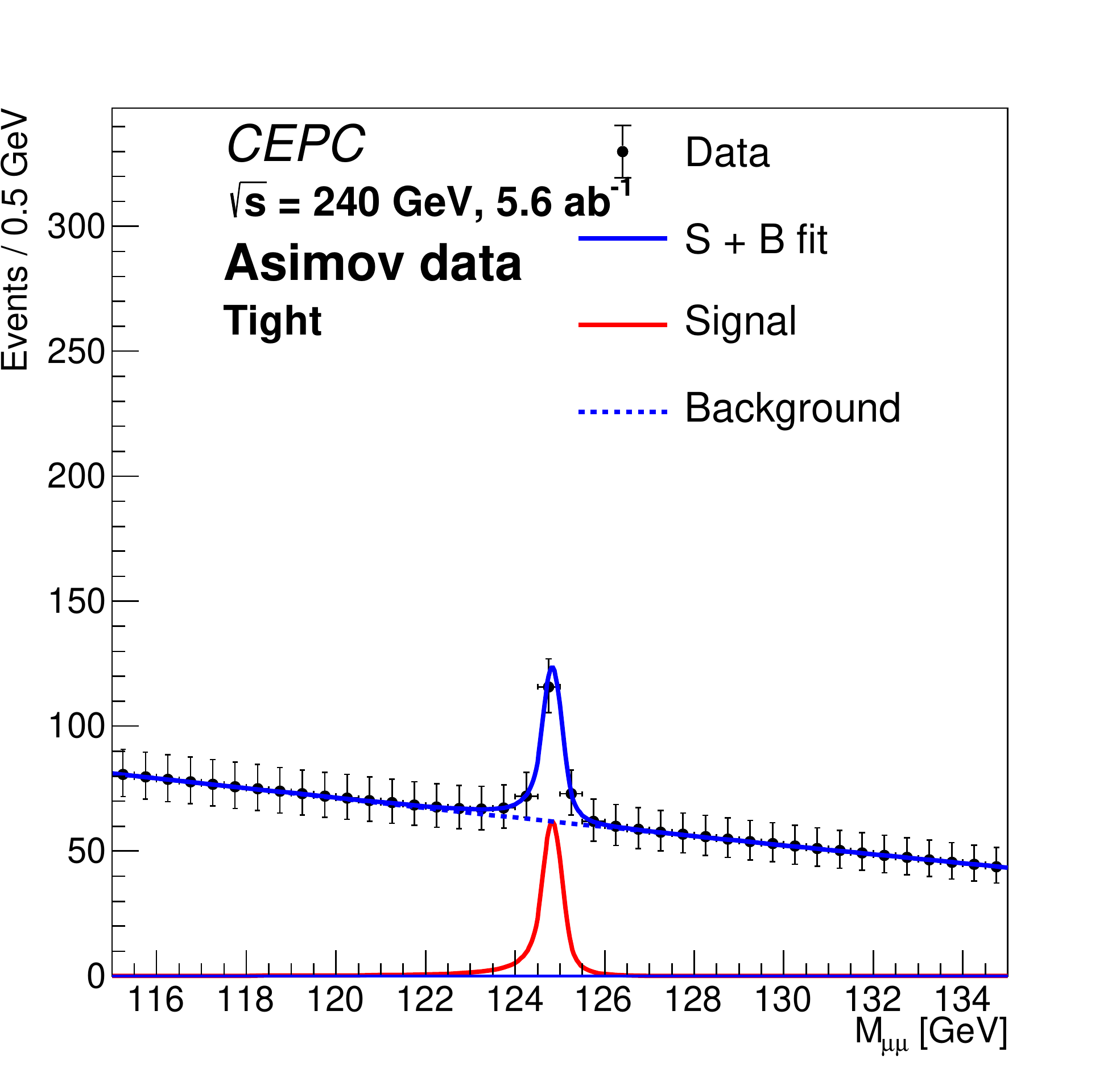}}
     \subfloat[Loose]{\includegraphics[width=0.5\linewidth]{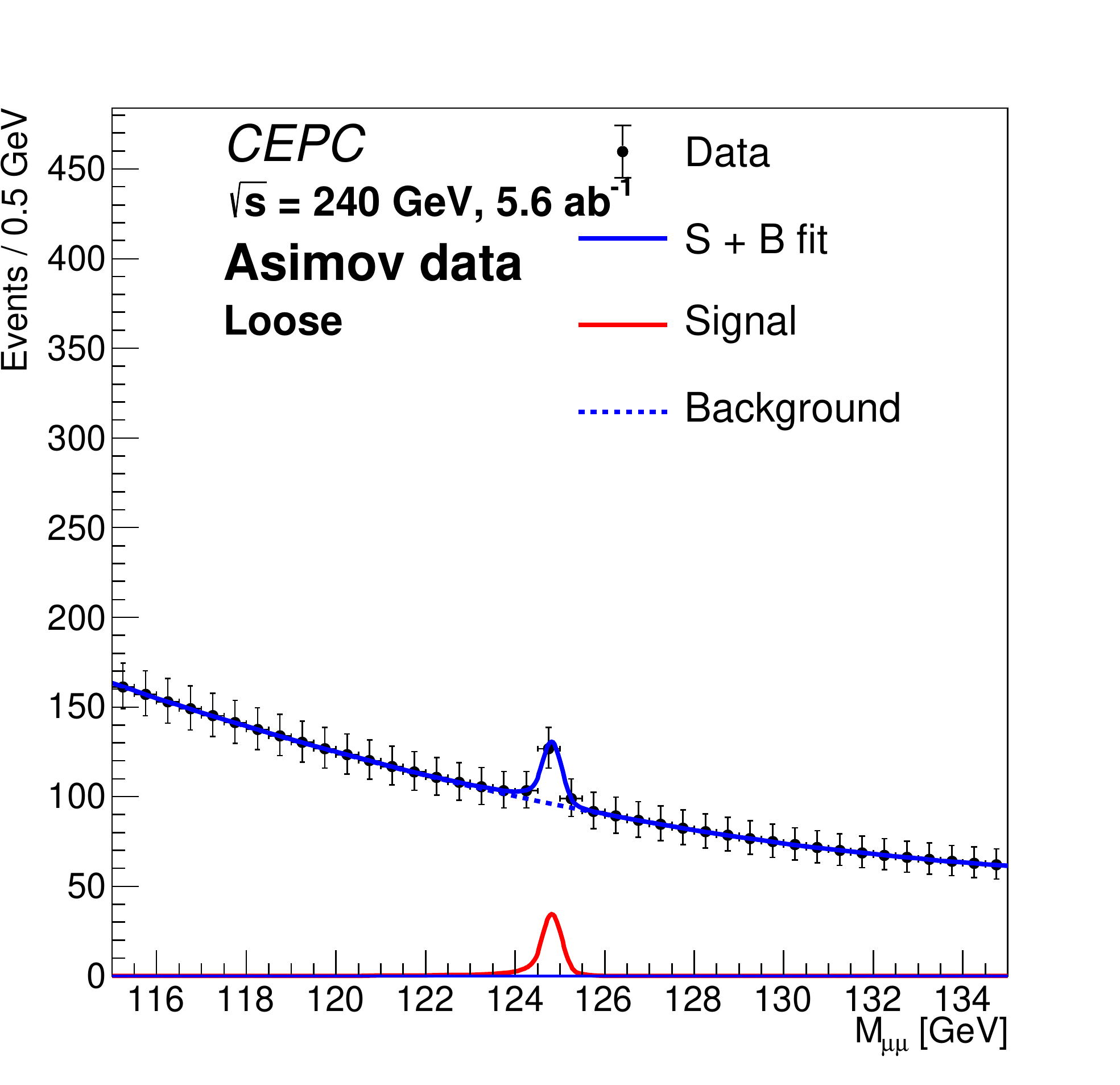}}
     \caption{The $m_{\mumu}$ distribution of the Asimov data and the fitted models in the tight (a) and the loose (b) categories. The blue curve is the fitted signal + background model. The red curve is the signal component and the dashed blue curve is the background component.}
     \label{fig:simultaneous}
\end{figure*}

In the analysis, to avoid the statistical fluctuations of the MC samples, the Asimov data~\cite{Cowan_2011} is generated and is fitted on to get the expected precision and significance of the signal process. Figure~\ref{fig:simultaneous} shows the $m_{\mumu}$ distribution of the Asimov data and the fitted models in two categories. The blue curve is the fitted signal + background model. The red curve is the signal component and the dashed blue curve is the background component. The expected signal
strength $\mu$ is estimated to be $1.00_{-0.18}^{+0.19}$ with statistical uncertainty. The corresponding significance is 6.1$\sigma$. In order to estimate the potential over-training effects on the background events during the BDTG event categorization (Section~\ref{BDTG}), the background categorization efficiency uncertainty accounting for the difference between the training and test background events is applied. The statistical uncertainties of the background model parameters ($a_{0}$, $a_{1}$
of the Formula~(\ref{eq:bkgModel})) calculated by fitting on the background MC samples are applied as well as the background shape uncertainties. It turns out above systematic impacts on the $H\to\mu^{+}\mu^{-}$ signal measurement precision and the significance are negligible, thus are neglected in the study. 

The High Luminosity Large Hadron Collider (HL-LHC) is an upgrade of the LHC which aims to collect $pp$ collision data with 3000 fb$^{-1}$ integrated luminosity at $\sqrt{s}=14$ TeV. The expected precision of the $H\to\mu\mu$ measurement in the ATLAS experiment is extrapolated from the analysis using 79.8 fb$^{-1}$ of data at $\sqrt{s}=13$ TeV~\cite{ATL-PHYS-PUB-2018-054}. Around 41k $pp\to H\to\mu\mu$ events will be generated at HL-LHC, and the precision in the extrapolation is estimated to be 14\%. While $\sim$167 $e^{+}e^{-}\to Z(\to q\bar{q})H(\to\mu\mu)$ events are expected to be generated at CEPC. With the help of the extremely high efficiency of muon events and clean backgrounds, the precision is at the same level between 2 analyses. The prospects for measuring the branching fraction of $H\to\mu\mu$ at the International Linear Collider (ILC) have been evaluated considering centre-of-mass energies ($\sqrt{s}$) of 250 GeV and 500 GeV~\cite{Kawada_2020}. For both $\sqrt{s}$ cases, two final states $e^{+}e^{-}\to q\bar{q}H$ and $e^{+}e^{-}\to \nu\bar{\nu}H$ have been analyzed. For integrated luminosities of 2 ab$^{-1}$ at $\sqrt{s}=250$ GeV and 4 ab$^{-1}$ at $\sqrt{s}=500$ GeV, both $ZH$ and $WW$ fusion production modes are considered and $\sim$199 signal events will be generated. The combined precision is estimated to be 17\%. In the Future Circular Collider electron-positron (FCC-ee) experiment~\cite{Mangano:2651294}, the expected uncertainty of $\sigma(e^{+}e^{-}\to ZH)\times BR(H\to\mu\mu)$ is measured by using 5 ab$^{-1}$ of data at $\sqrt{s}=240$ GeV. The precision of 19\% is compatible with the result estimated in the CEPC experiment.

\section{Discussions on the detector performance}
\label{Discussion}

\begin{figure*}[htbp]
\centering
        \subfloat[]{\includegraphics[width=0.5\linewidth]{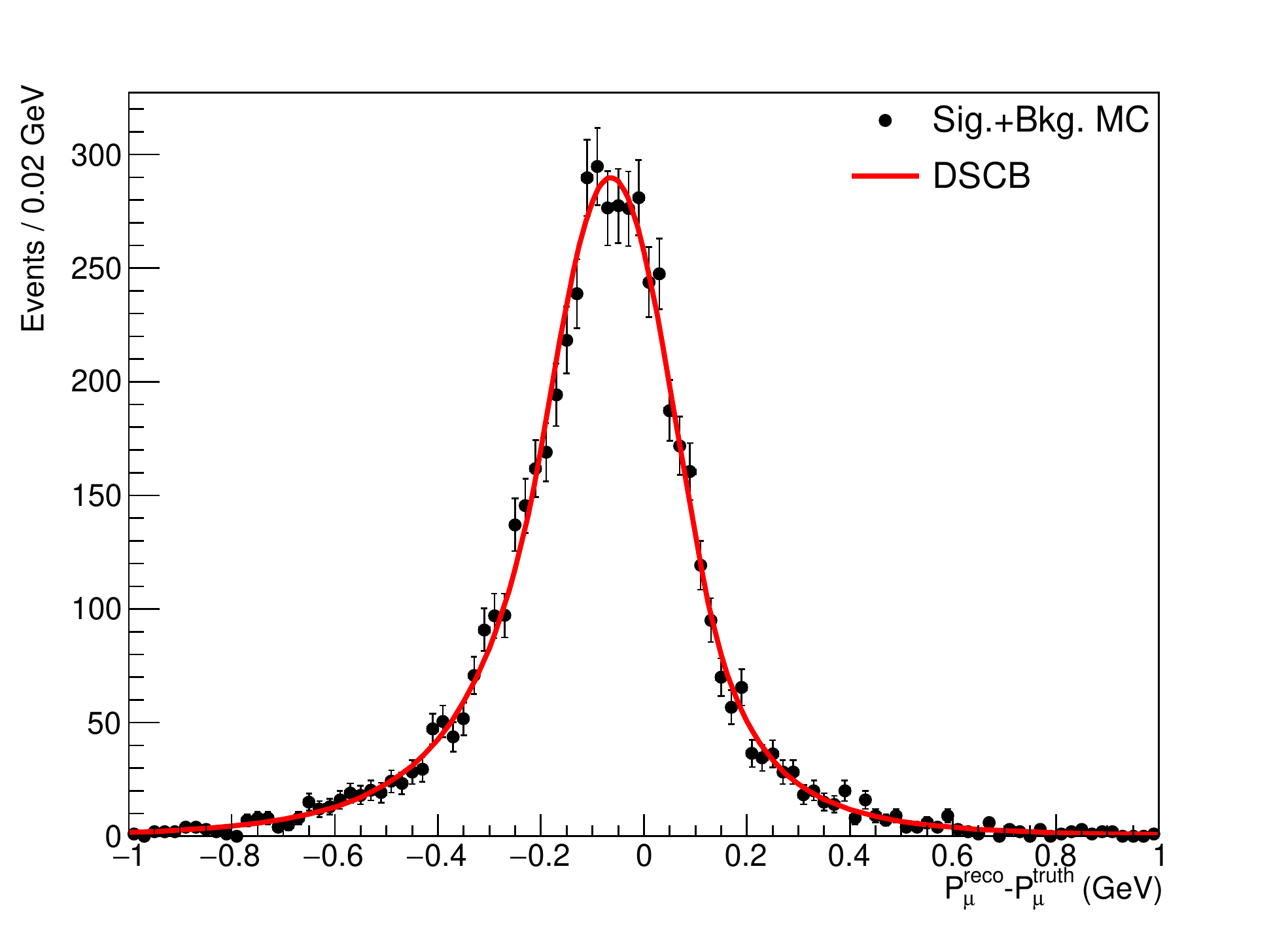}\label{fig:resolution}}
     \subfloat[]{\includegraphics[width=0.45\linewidth]{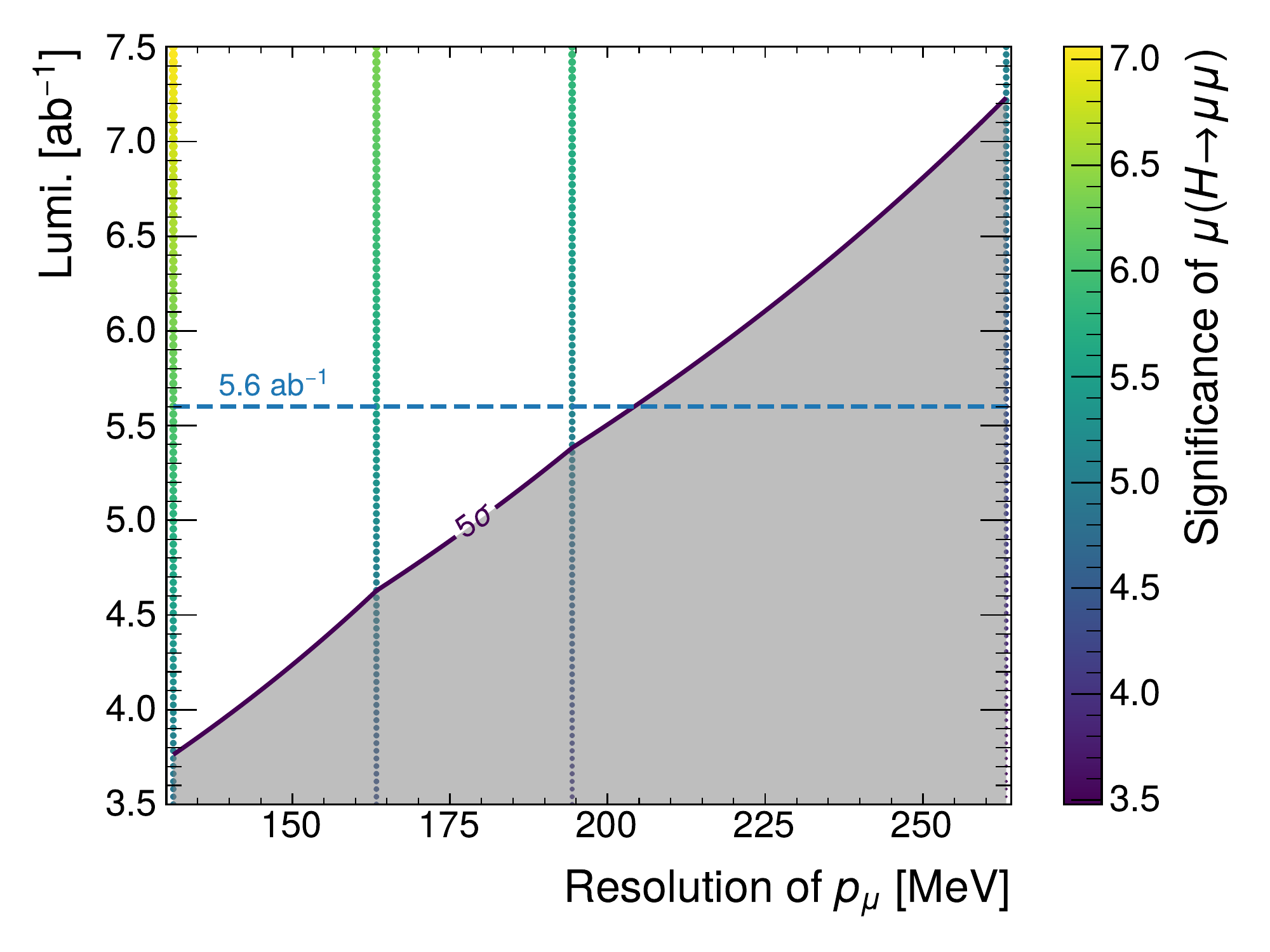}\label{fig:lumiVSRes}}
        \caption{(a) Nominal momentum resolution of the muon with the signal MC events passing all selections. The DSCB function is used to fit the spectrum and $\sigma_{CB}$ is measured to be 131 MeV. (b) Two dimensional expected significances of the $H\to\mu\mu$ process as a function of the integrated luminosity and the momentum resolution of the
    muon. Colored scatters are expected significances. For significances with the same momentum resolution, besides the measured numbers (Table~\ref{tab:asimov}) at the nominal integrated luminosity (5.6 ab$^{-1}$), others are scaled by $\sqrt{\frac{\mathcal{L}}{\mathcal{L}_{0}}}$, where $\mathcal{L}$ is the target integrated luminosity and $\mathcal{L}_{0}$ is the nominal one. It's assumed that the significance is only restricted by the number of events. The discovery curve
is extrapolated with points in the (resolution, integrated luminosity) space and the expected significances in the gray band are below 5$\sigma$.}
\end{figure*}

    In order to study the CEPC detector performance on muon measurements, the resolution of the muon momentum ($\sigma_{\mu}=(p_{\mu}^{reco}-p_{\mu}^{truth})$) is smeared by 25\%, 50\% and 100\%. The $H\to\mumu$ measurement is repeated to estimate the reduction in the signal precision. In the analysis, the nominal momentum resolution of the muon is performed in the Figure~\ref{fig:resolution} with the MC events (signal + background) passing all selections. The DSCB function is
    used to fit the spectrum and $\sigma_{CB}$ is measured to be 131 MeV.       

Table~\ref{tab:asimov} shows the expected signal strengths $\mu$, significances and the reductions in significances by smearing the resolution of the muon momentum. Figure~\ref{fig:lumiVSRes} shows the expected significances of the $H\to\mu\mu$ process in the 2 dimensional map of the integrated luminosity and the momentum resolution of the
muon. Colored scatters are expected significances. For significances with the same momentum resolution, besides the measured numbers (Table~\ref{tab:asimov}) at the nominal integrated luminosity (5.6 ab$^{-1}$), others are scaled by $\sqrt{\frac{\mathcal{L}}{\mathcal{L}_{0}}}$, where $\mathcal{L}$ is the target integrated luminosity and $\mathcal{L}_{0}$ is the nominal one. It's assumed that the significance is only restricted by the number of events. The discovery curve
is extrapolated with points in the (resolution, integrated luminosity) space and the expected significances in the gray band are below 5$\sigma$. The resolution has to be better than 204 MeV to discover the $H\to\mu\mu$ process at the nominal integrated luminosity. With the nominal muon momentum resolution of the detector, the integrated luminosity should be greater than 3.8 ab$^{-1}$ for the discovery of the di-muon process. In the worst case that the resolution is 100\% worse than the designed parameters, the integrated luminosity should be greater than 7.2 ab$^{-1}$.

\begin{table}[H]
    \small
    \centering
\tabcaption{The expected signal strength $\mu$, significance and the reduction in significance with the resolution of the muon momentum smeared by 25\%, 50\% and 100\%.}
\begin{tabular}{cccc}
     \hline
    Smearing& 25\% & 50\%  &100\%  \\ \hline
	$\mu$ & $1.00_{-0.20}^{+0.21}$& $1.00_{-0.21}^{+0.22}$& $1.00_{-0.24}^{+0.25}$ \\
	Significance& $5.5 \sigma$& $5.1 \sigma$& $4.4 \sigma$ \\
	Reduction in significance &10\%& 16\%& 28\% \\ \hline
\end{tabular}
\label{tab:asimov}
\end{table}
%

\section{Conclusion}
\label{Conclusion}

The $e^{+}e^{-} \rightarrow Z(\to q\bar{q})H(\to\mu^{+}\mu^{-})$ process is studied using Monte-Carlo events in the CEPC experiment. The simulated samples are generated with the center of mass energy of 240 GeV. The event selections are updated and the categorization is optimized by the BDTG method to improve the signal significance. The maximum unbinned likelihood fit method is applied to fit on the Asimov $m_{\mumu}$ distributions in two event categories simultaneously. With the designed integrated luminosity of 5.6
ab$^{-1}$, the statistical-only precision of the expected signal is 19\% and the corresponding significance is 6.1$\sigma$. The systematic uncertainties  from the background MC statistical fluctuations are evaluated and are tested to have negligible impacts on the $H\to\mu^{+}\mu^{-}$ signal. The performance of the CEPC detector is further studied by smearing the resolution of the muon momentum by 25\%, 50\% and 100\% in the simulated Monte Carlo samples. The impacts on the signal precision of this analysis are estimated. The resolution has to be better than 204 MeV to discover the $H\to\mu\mu$ process at the nominal integrated luminosity. If the resolution is 100\% worse than the designed parameters, the integrated luminosity should be greater than 7.2 ab$^{-1}$ for the discovery.

\section{Acknowledgments}
\acknowledgments{We acknowledge the support of the Innovative Research Program of IHEP (E2545AU210); CAS Center for Excellence in Particle Physics; Yifang Wang’s Science Studio of the Ten Thousand Talents Project; the CAS/SAFEA International Partnership Program for Creative Research Teams (H751S018S5); IHEP Innovation Grant (Y4545170Y2); Key Research Program of Frontier Sciences, CAS (XQYZDY-SSW-SLH002); Chinese Academy of Science Special Grant for Large Scientific Project (113111KYSB20170005);
the National Natural Science Foundation of China (11675202); the National 1000 Talents Program of China; Fermi Research Alliance, LLC (DE-AC02-07CH11359); the NSF(PHY1620074); the Maryland Center for Fundamental Physics (MCFP); Tsinghua University Initiative Scientific Research Program; and the Beijing Municipal Science and Technology Commission project (Z181100004218003).}

\vspace{10pt}
\bibliographystyle{apsrev}
\bibliography{references}

\end{multicols}

%
%
%
%
%
%

\end{document}